\documentclass[useAMS,usenatbib]{mn2e}

\def\aap{AA}

\def\apjl{ApJL}

\def\mnras{MNRAS}
\def\apj{ApJ}
\def\apjs{ApJS}
\def\aj{AJ}

\def\memsai{MmSAI}

\usepackage{graphicx}
\usepackage{float}
\usepackage{amssymb}
\usepackage{amsfonts}
\usepackage{amsmath} 
\usepackage{color}

\def\ucsb{Department of Physics, University of California, Santa Barbara, CA 93106, USA}
\def\ucla{Physics and Astronomy Building, 430 Portola Plaza, Box 951547, Los Angeles, CA 90095-1547, USA}
\def\kipac{Kavli Institute for Particle Astrophysics and Cosmology, Stanford University, 452 Lomita Mall, Stanford, CA 94305, USA}

\def\aaemail{\tt aagnello@physics.ucsb.edu, tt@astro.ucla.edu}

\title[Mining for lensed QSOs]{Data Mining for Gravitationally Lensed Quasars} 
\author[Agnello et al.]{
  Adriano Agnello$^{1,2}$\thanks{\aaemail},
  Brandon C. Kelly$^1$,
  Tommaso Treu$^{1,2}$,
  and Philip J. Marshall$^{3}$
  \medskip\\
  $^1$\ucsb\\
  $^2$\ucla\\
  $^3$\kipac\\
}

\begin{document}

\voffset-.6in

\date{Accepted . Received }

\pagerange{\pageref{firstpage}--\pageref{lastpage}} 

\maketitle

\label{firstpage}

\begin{abstract}
Gravitationally lensed (GL) quasars are brighter than their unlensed
counterparts and produce images with distinctive morphological
signatures. Past searches and target selection algorithms, in
particular the Sloan Quasar Lens Search (SQLS), have relied on
basic morphological criteria, which were applied to samples of bright,
spectroscopically confirmed quasars.  The SQLS techniques are not
sufficient for searching into new surveys (e.g. DES, PS1, LSST),
because spectroscopic information is not readily available and the
large data volume requires higher purity in target/candidate
selection. 
%Conversely, modeling-based automatic lens finding algorithms in
%imaging datasets require
%large amounts of CPU and scientist time that seem prohibitive for
%upcoming surveys. To overcome these limitations
We carry out a
systematic exploration of machine learning techniques and demonstrate
that a two step strategy can be highly effective. In the first step we
use catalog-level information ($griz$+WISE magnitudes, second moments)
to preselect targets, using artificial neural networks.  The accepted
targets are then inspected with pixel-by-pixel pattern recognition
algorithms (Gradient-Boosted Trees), to form a final set of
candidates. 
% Both methods are tested against the SQLS sample and
%suitably interpreted in terms of selection bias, thereby obtaining a
%robust characterization of candidates selected this way. 

 The results from this procedure can be used to further refine the simpler SQLS
algorithms, with a twofold (or threefold) gain in purity and the same
(or $80\%$) completeness at target-selection stage, or a purity of
$70\%$ and a completeness of $60\%$ after the candidate-selection step.
% Our catalog level selection alone is significantly more effective
%than simple color cuts. For comparison,
Simpler photometric searches in $griz$+WISE based on colour cuts
 would provide samples with $7\%$ purity or less. Our technique is extremely fast, as
a list of candidates can be obtained from a stage III
experiment (e.g. DES catalog/database) in {a few} CPU hours.
 The techniqus are easily extendable to Stage IV experiments like LSST with the addition
of time domain information.

\end{abstract}

\section{Introduction}

Gravitationally lensed quasars are a very useful astrophysical tool. They can
be used to investigate a variety of phenomena \citep[e.g.][]{cou02,jac13} --
often providing unique insights -- including cosmography
\citep[e.g.][]{suy14}, the free streaming length of dark matter
\citep{nie14,mao98,met01,dal02,roz06}, the properties of quasar host galaxies
\citep[e.g.][]{pen06}, the dust extinction law in distant galaxies
\citep[e.g.][]{dai06}, the stellar initial mass function
\citep[e.g.][]{ogu14}, the size of accretion disks \citep[e.g.][]{bla14} and
the structure of the broad line region \citep{gue13,slu12}.

Unfortunately, lensed quasars are extremely rare on the sky, as the phenomenon
requires the alignment of a deflector and a source, typically within
arcseconds. The occurrence of a strongly lensed quasar depends on the optical
depth of deflectors (typically massive galaxies or groups), and the density of
quasars on the sky. At the typical depth of current ground-based wide field
surveys, the abundance of lensed quasars is approximately 0.1 per square
degree \citep{om10}. Most lensed quasars are expected to be doubly imaged,
with quadruply imaged systems comprising approximately one sixth of the total,
because the inner caustics are typically significantly smaller than the outer
ones. The areas surveyed to date have led to a current sample comprising of
order only a hundred lensed quasars. Given that each specific application is
usually best suited for limited subsamples (e.g. quadruply imaged, or highly
variable, or radio-loud sources), most of the analyses so far have been
limited to samples of 10-20 objects at best. In the vast majority of cases,
sample size is the main limiting factor of present day studies.
% Adjust the bibliography in introfirst

Systematic searches of lensed quasars in the optical have been carried
out in the Sloan Digital Sky Survey (SDSS)\footnote{Surveys referred
to here: SDSS, Sloan Digital Sky Survey \citep{yor00}; PS1, the first
of Panoramic Survey Telescope and Rapid Response System (Pan-STARRS)
telescopes, {http://pan-starrs.ifa.hawaii.edu/public/}; DES, Dark
Energy Survey \citep{san10} ; Gaia \citep{per01}; LSST, Large Synoptic
Survey Telescope, \citep{ive08}; HSC, Hyper-Suprime Cam,
{http://subarutelescope.org/Observing/Instruments/HSC/index.html}; and
WISE, the Wide-field InfraredSurvey Explorer \citep{wri10}.}.
\citet{pin03} examined objects in the Early Data Release, flagged as
quasars based on spectroscopic criteria, whose image cutouts showed
evidence of multiple sources, parameterized by the best-fit
$\chi^{2}.$ The SDSS Quasar Lens Search \citep[hereafter
SQLS,][]{ogu06} extended that approach, exploiting the information
already available at catalogue level before cutouts were inspected,
with a strategy tailored to two different regimes. The search for
systems with large (approx.$\geq3''$) image separation selected
spectroscopically confirmed quasars with nearby companions with
similar colours ({colour} selection).  For small-separation
systems, which are not succesfully deblended by the SDSS pipeline, the
algorithm selects spectroscopically-confirmed quasars with an extended
morphology, signaled by a low stellarity likelihood in the $ugri$
bands ({morphological} selection).  This procedure produced a
valuable sample of lensed quasars brighter than 19.1 in $i-$band by
the SDSS seventh Data Release \citep[DR7][]{aba09}, 26 of which have
well defined population properties \citep{ina12}.  In particular, out
of 54 morphologically-selected candidates, 10 were true
small-separation lensed quasars.

New or upcoming surveys will deliver a wealth of new systems, thanks
to improved depth and larger footprint. Oguri and Marshall (2010,
hereafter OM10) have predicted the distribution of strongly lensed
QSOs, providing estimates of the abundance of these systems as a
function of survey depth. Similar (although somewhat simpler)
estimates have been made {by \citet{fin12}} for the Gaia space
mission, adopting a $G-$band limiting magnitude of 20. The results are
summarised in Table~\ref{tab:numbers}.  While the total numbers can
vary among different surveys, in general we can expect one lensed QSO
every five square degrees at an $i-$band depth of 24. Most of the
lensed QSOs will be doubly imaged, while about a sixth of the
population consists of highly informative quad configurations (OM10).
Approximately a tenth of the expected systems are brighter than 21 in
$i-$band.
 
Given these numbers, sharp techniques are required in order to obtain
a sample of targets with sufficient purity to enable efficient
follow-up.  If the ratios of true and false positives seen in SQLS
were to hold for new surveys as well, this strategy would soon become
unfeasible in surveys deeper or wider than SDSS. In fact, the false
positive rate is likely to increase in new surveys, due to the lack of
ready spectroscopic confirmation for QSO-like objects.

Fortunately, some aspects of the SQLS strategy can be improved. For
example, survey catalogues contain more morphological information
(i.e. second moments, axis ratio and position angle) besides the mere
stellarity likelihood. In principle, this can be used to skim the
catalogue for {targets}, without significant slow-down. Once
the targets are selected, their image cutouts can be examined with
pixel-by-pixel pattern recognition techniques, which mimic the common
practice of selecting candidates (or rejecting obvious outliers)
through eyeballing. The final result is a pool of {candidates},
which can then be followed up with better imaging. 

{Data mining} is the process of uncovering relations between
observables, and therefore isolating relevant information, from large
samples of objects. 
% PJM: need to add in here some introduction of how catalog mining
% can be done, incl citations. Then move on to introduce image
% classification. At the moment they are sort of jumbled together.
In particular, from the viewpoint of the lensed
QSO search, pattern recognition algorithms help isolate the promising
targets and candidates. Similar machine learning approaches have been
followed in other areas of astrophysics, such as variable stars and
transients \citep{bel03,bla14}, galaxy classification \citep{kel05} or
in general object classification and photometric redshift
\citep{bal10, car14} for SDSS objects, as well as supernova lightcurve
classification \citep[and references therein]{ish13}, but not yet to
the search for lensed quasars. Here we illustrate a first step in this
direction. 
% PJM: maybe here is a good place to discuss lens finding as a
% classification problem, which means, if you go for supervised methods,
% you have to define classes and start thinking about a training set. I
% think both these things need introducing so that the ``Data'' section
% makes sense.}
We note that variability provides additional information
which might be very effective in identifying lensed quasars 
\citep[e.g.][]{Pindor05,Kochanek++06}. We do not include this
information in this first exploratory study; however, our procedure is
easily generalizable to multi-epoch data in order to take advantage of
this additional feature for selection.

In order to assess the performance of machine learning in this area,
we examine the problem of finding strongly lensed quasars in SDSS,
focussing on the small-separation regime, when the multiple images and
the deflector are expected to be highly blended. This is the regime
where we expect most of the candidates to be found \citep{om10}, and
also the one that is most challenging, from a conceptual and
computational viewpoint, since both the QSO images and the galaxy are
blended together. Of course, our data mining approach can easily be
extended to systems with larger image-separation; we discussthis
briefly in Section \ref{sect:deblended}.

This paper is structured as follows.  In Section 2 we describe the
data sets used for training, validating and testing our machinery.  In
Section 3 we introduce the techniques used in this work, leaving a
more detailed discussion in the Appendices for the interested reader.
 Section 4 shows the results of target- and
candidate-selection on simulated data, with an application to the SQLS
sample of morphologically selected targets from {SDSS DR7
\citep{ina12}.} Extensions to the deblended regime are illustrated
 in Section \ref{sect:deblended}.
 Finally, we conclude in Section 6.
\begin{table} 
\centering
\begin{tabular}{|c|c||c|c|}
\hline
survey & depth & lensed & unlensed\\
\hline
DES & 24.0 & 0.23 & 740\\
PS1 & 22.7 & 0.07 & 250\\
Gaia & 20.0 & 0.06 & 12.5\\
LSST and HSC & 24.9 & 0.4 & 1175 \\
%HSC & 24.9 & 0.41 & 1173 \\
%Skymapper & ... & ... & ...\\
\hline
\end{tabular}
\caption{Number of lensed and unlensed QSOs per square degree in new
 or upcoming surveys, adapted {from \citet{om10} and \citet{fin12}}.
 The depth is in $G-$band for Gaia and $i-$band for the other
 surveys.}
\label{tab:numbers}
\end{table}
\begin{figure*}
 \centering
 \includegraphics[width=0.33\textwidth]{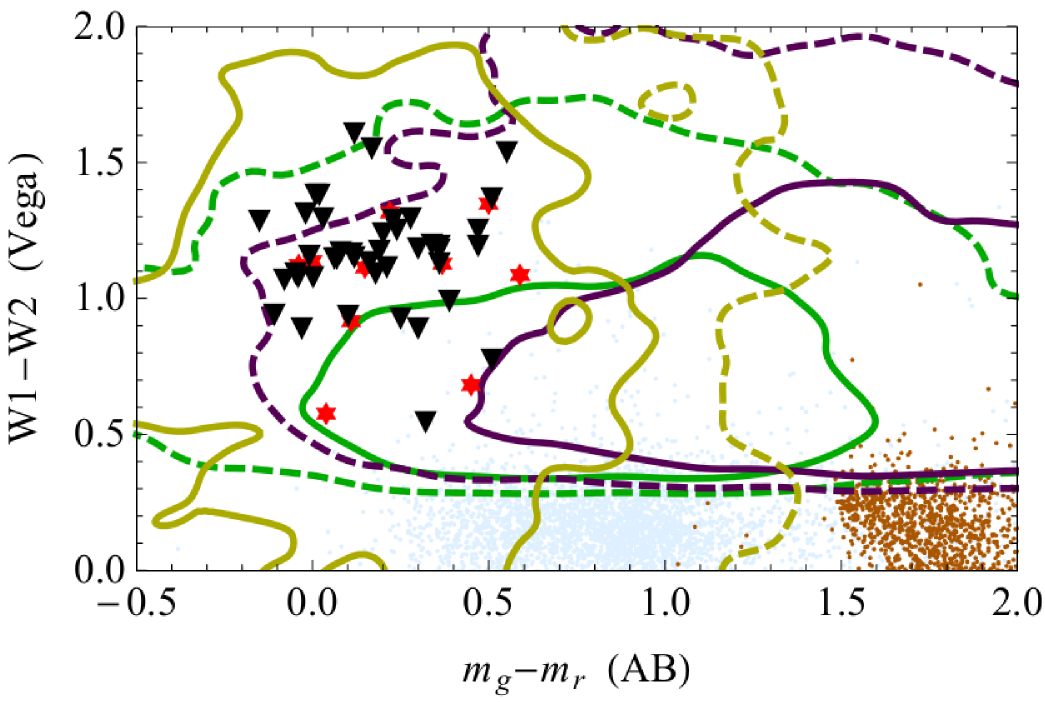}
 \includegraphics[width=0.33\textwidth]{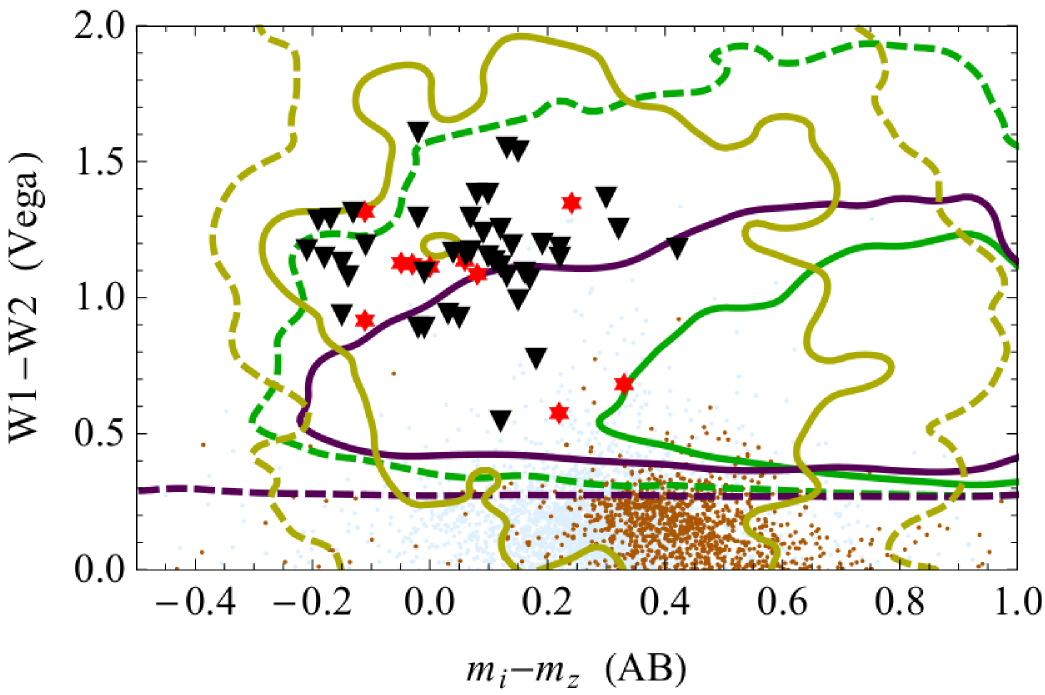}
 \includegraphics[width=0.33\textwidth]{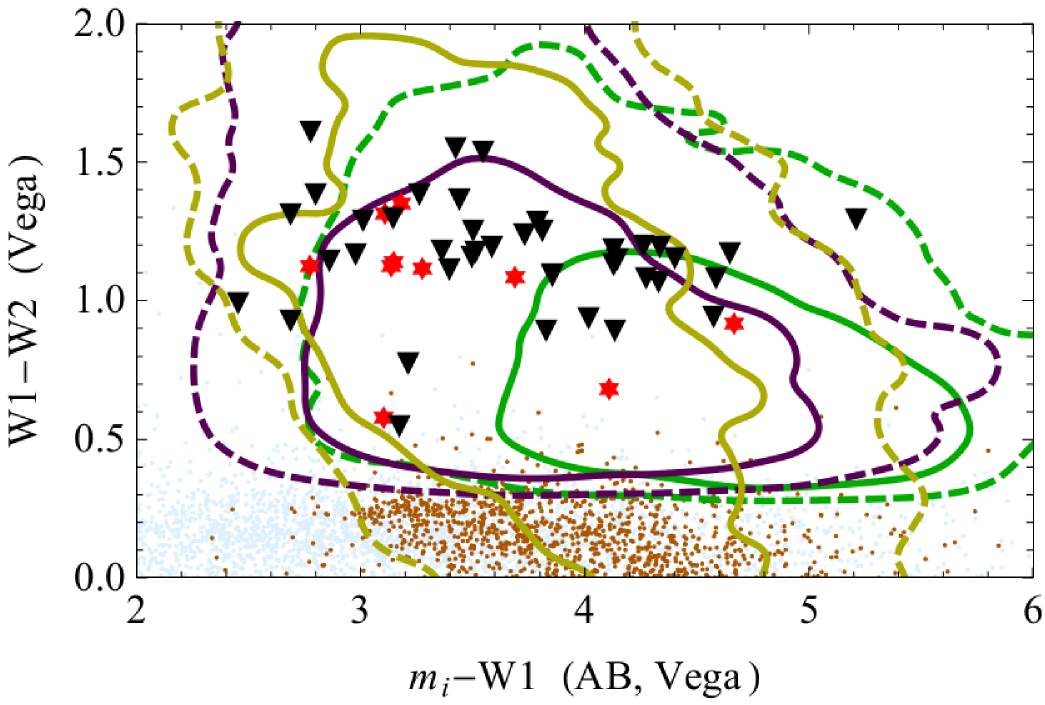}\\
 \includegraphics[width=0.33\textwidth]{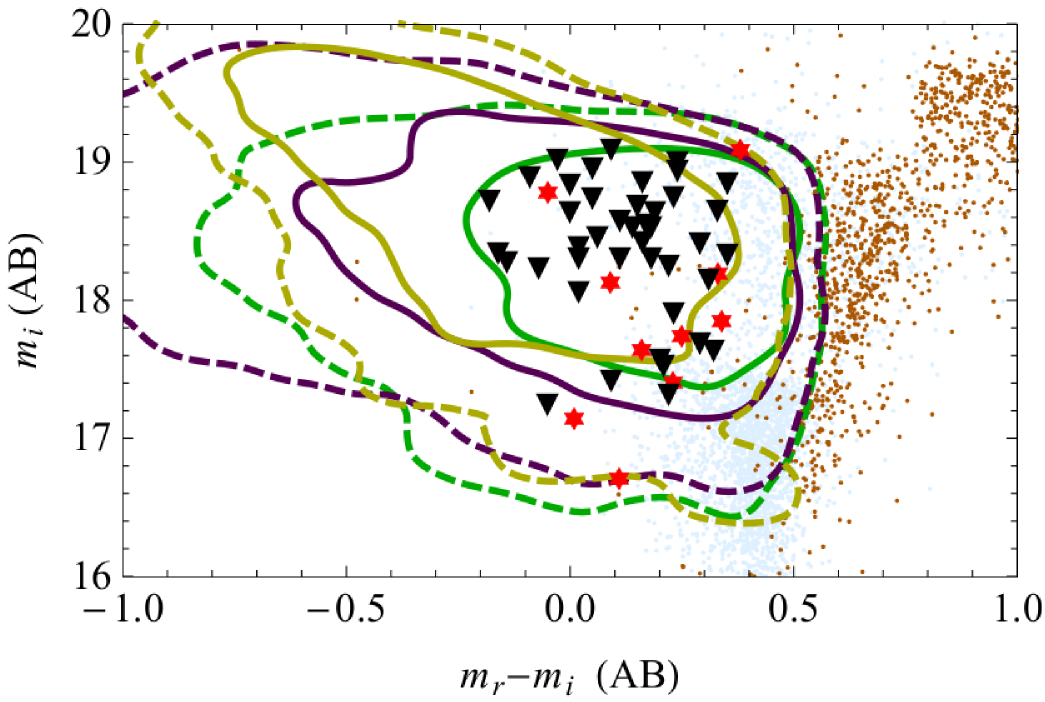}
 \includegraphics[width=0.33\textwidth]{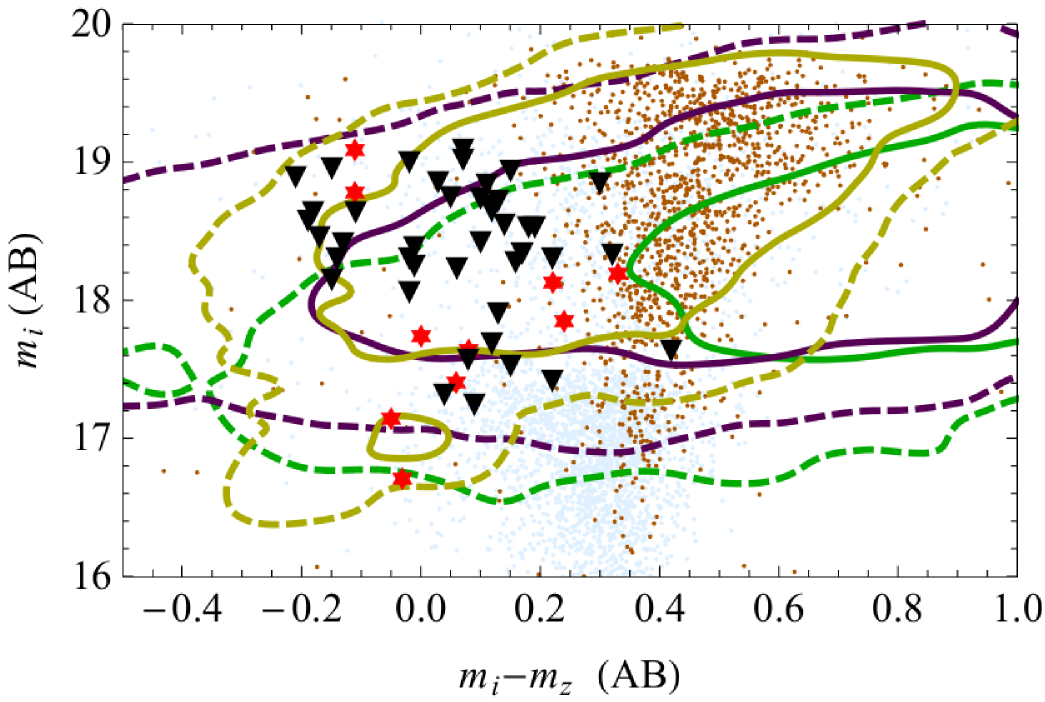}
 \includegraphics[width=0.33\textwidth]{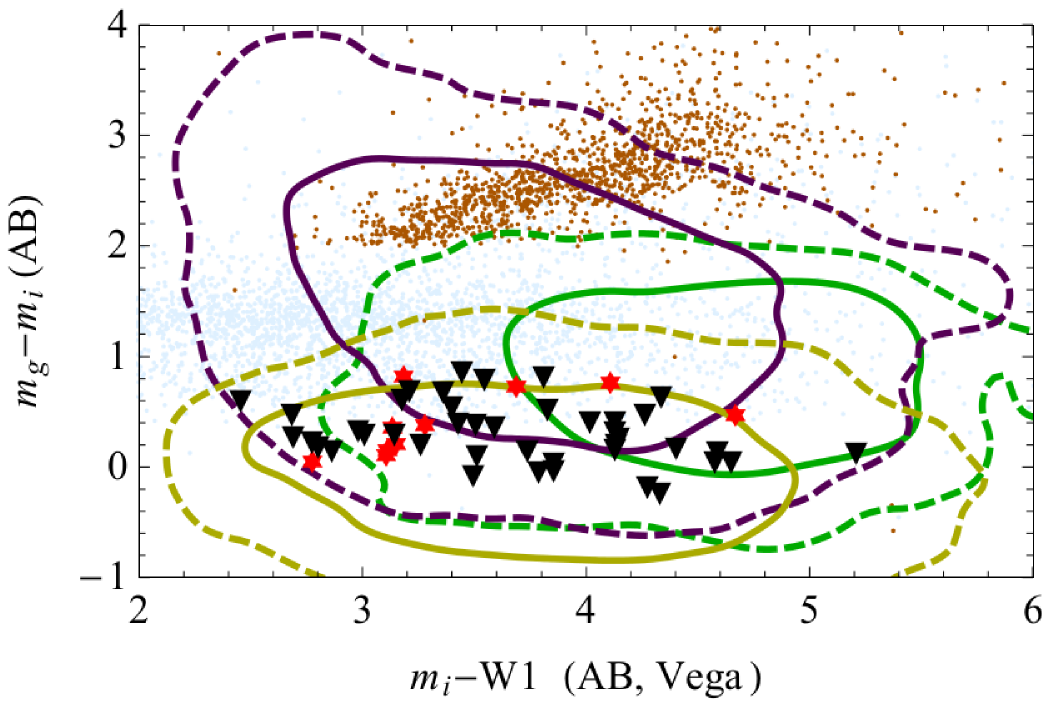}\\
\caption{\small{Colour-colour plots for some objects of interest in this search. Green (resp. purple, yellow) contours
 delimit the $68\%$ and $95\%$ of lensed QSO (resp.unlensed QSO plus LRG, QSO alignment) populations, with blended images.
 Orange (resp. light blue) dots mark the Luminous Red and Blue Cloud galaxy populations. Red stars (resp. black triangles)
 mark the true (resp. false) positives in the SQLS morphologically selected sample from {SDSS DR7 \citep{ina12}}. The
 candidates cover predominantly the locus of QSO alignments, which is also contaminated by B.C.galaxies even when the WISE
 bands are used.}}
\label{fig:colcolplots}
\end{figure*}
\section{Data}
\label{sect:data}
Our aim is to obtain a partition of objects into different {classes}, one of which will be the truely lensed QSOs.
Hence, we must ensure that the selection algorithms are accurate enough to distinguish between different classes.
 To do so, we deploy a mixture og real and simulated systems, so as to build large enough samples to train, validate and test hte algorithms, anticipating the number of lensed quasars and false
positives found in past searches not to be sufficient to the purpose.
 The data mining details are described in the next Section, here we give an overview of the samples used in this work.

\subsection{Data from SDSS and WISE}
\label{sect:realobjects}

\begin{table} 
\centering
\begin{tabular}{|c|c|c|}
\hline
\multicolumn{3}{c}{SDSS imaging conditions} \\
\hline
band & sky & PSF FWHM\\
 & $(ABmag/arcsec^{2})$ & $arcsec$ \\
\hline
 $g$ & $21.9\pm0.3$ & $1.65\pm0.4$ \\
 $r$ & $20.9\pm0.3$ & $1.4\pm0.3$ \\
 $i$ & $20.2\pm0.4$ & $1.4\pm0.3$ \\
 $z$ & $18.9\pm0.5$ & $1.4\pm0.3$ \\
\hline
\end{tabular}
\caption{Imaging conditions for SDSS simulated objects. We list the
 mean value and standard deviation of sky brightness (magnitudes per
 square arcsecond) and image quality (PSF FWHM) in the four $griz$
 bands. SDSS image quality is typically worse in the bluest ($g$)
 band, which is also where the sky is typically fainter.}
\label{tab:SDSSimg}
\end{table}

In order to compare our methods with past searches, in particular the
SQLS, we investigate lensed quasar detection in SDSS-like imaging
conditions, which are summarized in Table~\ref{tab:SDSSimg}. For the
target-selection step, the photometry is given by SDSS $griz$ bands,
plus the infrared bands $W1$ and $W2$ from WISE. We do not make use of
the SDSS $u$ band, because this is not always available in upcoming
surveys (like DES and PS1). The morphological parameters (axis ratios,
p.a.s) are computed from $25\times25$-pixel simulated cutout images 
in $griz$ bands, which are produced as described in
Section~\ref{sect:simpop} and Appendix A below.  WISE has a PSF with
FWHM$\approx6'',$ which makes it of limited use for evaluating
morphologies. For the candidate-selection step, we consider just the
cutouts in $griz$ bands, without additional information from WISE
photometry.

\subsection{Object Classes}
\label{sect:classes}

We use a classification scheme resembling the outcome of the SQLS. For
the target selection, a simulated system 
can be: a lensed quasar, with
a Luminous Red Galaxy (LRG) as the deflector (labelled
{l.QSO}); a pair of closely aligned quasars, with different
redshifts ({2QSO}); or an alignment of LRG and unlensed quasar
({QSO+LRG}).

 Due to the absence of a spectroscopic selection
stage, we add as a contaminant a class of Blue Cloud galaxies
({BC}), with observables taken directly from SDSS.
The queries for Blue-Cloud and Luminous-Red galaxies are adapted to our
needs from a publicly available version on the SDSS website.
The BC class is also useful for dealing with objects with a strong
stellar component, e.g. an alignment of QSO and nearby star, that
would be harder to properly simulate.

 Figure~\ref{fig:colcolplots}
shows these systems in colour-colour space. The need for WISE
photometry is evident from the overlap of BC objects with our targets
of interest. Also, simple two-colour cuts in SDSS/WISE bands cannot
prevent the leakage of BC objects in the quasar locus, whence the
necessity of considering nontrivial combinations of bands, which are
naturally selected by data mining algorithms. 

\subsection{Simulated Populations}
\label{sect:simpop}

\begin{figure*}
 \centering
 \includegraphics[width=0.475\textwidth]{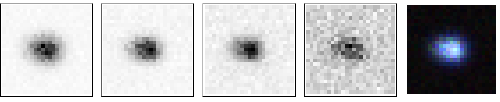}
 \includegraphics[width=0.475\textwidth]{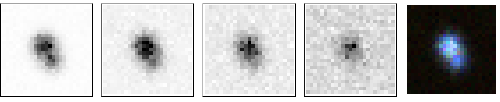}\\
 \includegraphics[width=0.475\textwidth]{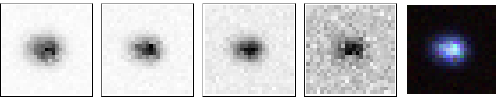}
 \includegraphics[width=0.475\textwidth]{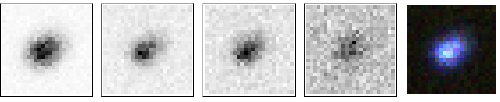}\\
 \includegraphics[width=0.475\textwidth]{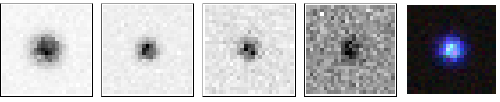}
 \includegraphics[width=0.475\textwidth]{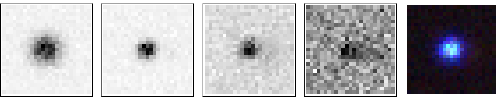}\\
 \includegraphics[width=0.475\textwidth]{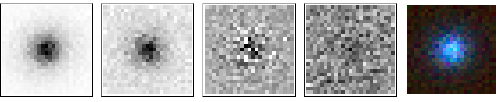}
 \includegraphics[width=0.475\textwidth]{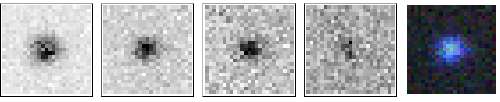}\\
 \includegraphics[width=0.475\textwidth]{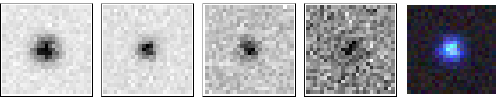}
 \includegraphics[width=0.475\textwidth]{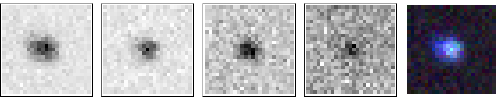}\\
 \includegraphics[width=0.475\textwidth]{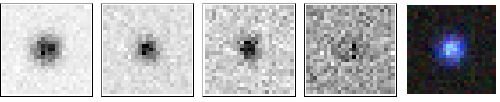}
 \includegraphics[width=0.475\textwidth]{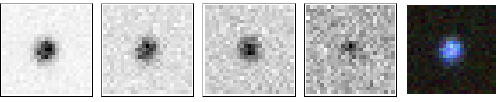}\\
 \includegraphics[width=0.475\textwidth]{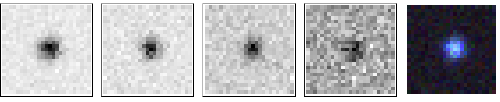}
 \includegraphics[width=0.475\textwidth]{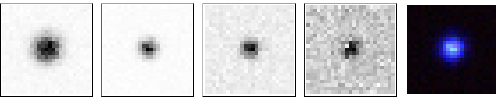}\\
 \includegraphics[width=0.475\textwidth]{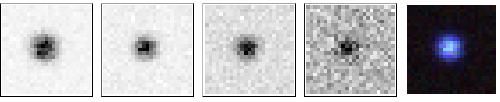}
 \includegraphics[width=0.475\textwidth]{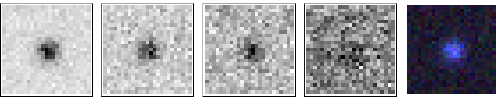}\\
\caption{\small{Example of the systems simulated in this work. Rows 1
and 2: lensed QSOs; rows 3 and 4: a QSO+LRG alignment, rows 5 and 6: a
QSO+QSO alignment; rows 7 and 8: a single (isolated, not-lensed) QSO.
Bands are $griz$ from left to right, the last sub-panel in each
sequence is a $gri$ composite.}}
\label{fig:puppies}
\end{figure*}
In order to reproduce the populations of systems that are expected in a survey, we
need to draw the quasars and galaxies from a distribution in redshift
and intrinsic properties.
We adhere to the common choices for QSO and LRG distributions, as
reviewed by OM10, who also used them to generate a mock population of
lensed quasars. In particular:
\begin{equation}
 n_{gal}(\sigma,z_{g})\mathrm{d}\sigma\mathrm{d}z_{g}\propto (\sigma/\sigma_{\star}(z_{g}))^{\alpha}\exp[-(\sigma/\sigma_{\star}(z_{g}))^{\beta}]\mathrm{d}z_{g}\mathrm{d}\sigma/\sigma
 \label{eq:lrg}
 \end{equation}
 \begin{equation}
 n_{qso}(m_{i,q},z_{q})\mathrm{d}m_{i,q}\mathrm{d}z_{q}\propto \frac{10^{-0.4(\alpha+1)(M_{i}-M_{\star}(z_{q}))}}{1+10^{0.4(\beta-\alpha)(M_{i}-M_{\star}(z_{q}))}}\mathrm{d}m_{i,q}\mathrm{d}z_{q}
 \label{eq:qso}
\end{equation}
(parameters further discussed in OM10), where $M_{i}$ is the $i-$band
magnitude that the QSO would have at $z_{q}=2$ as computed by
\citet{ric06}. For the population of lensed quasars, we directly use
the mock catalogue created by OM10. This provides, for each system:
the redshift $z_g,$ velocity dispersion $\sigma_g,$ axis ratio and
position angle of the lensing galaxy; and the redshift $z_{q},$
unlensed $i-$band magnitude $m_{i,q},$ image positions (relative to
the lens) and magnifications of the source quasar. The observables $z_{g},z_{q},\sigma,m_{i}$
 and their distributions (eq.~\ref{eq:lrg},~\ref{eq:qso}) can be used also for the $QSO+LRG$
 and $2QSO$ simulated classes.
 
When generating mock observations of these systems, one needs a
procedure to prescribe fluxes in different bands, for both the QSO and
LRG, and the effective radius of the LRG. The {primary}
observables $\mathbf{p}=(z_{q},z_{g},m_{i,q},\sigma_g)$ are at our
disposal, either drawn from equations (\ref{eq:lrg},\ref{eq:qso}) or
from the mock sample of OM10. The remaining observables $\mathbf{r}$
must be matched to those, taking account of intrinsic scatter in those
properties \citep[see][for the role of scatter in population
properties]{mit05}. In other words, if the entries of $\mathbf{p}$ are
given, then $\mathbf{r}$ must be drawn from a conditional distribution
$\theta(\mathbf{r}|\mathbf{p}).$
The detailed procedure is described in the Appendix, here we summarize
its main steps. First, we assemble a sample of LRGs and QSOs, with all
the relevant observables,  from SDSS and WISE. Then, once values of
$\mathbf{p}$ are assigned, a sparse interpolation procedure allows us
to build a smooth $\theta(\mathbf{r}|\mathbf{p})$ from the objects
within the sample in the vicinity of $\mathbf{p}.$ This way, every
time the primary observables are assigned, we can properly draw the
other observables within the whole range of LRGs that
match $z_g$ and $\sigma$, and QSOs that match $z_q$ and $m_i$.

Within the simulated data sets, we retain just those systems that are
brighter than the survey limit ($i-$band magnitude brighter than 21)
and with signal-to-noise ratio at least $5,$ which helps prune extreme
fluctuations in the simulated sky noise. The cutouts are not
necessarily aligned with the p.a. of the image, which forces the
learners to concentrate just on those features that are intrinsic to
the systems and physically relevant. For the same reason, they are
slightly off-centered in different bands -- by at most two pixels, a
common situation that occurs when downloading image cutouts. We
emphasize that the populations simulated here are more general than
the SQLS dataset, which relied on a heavy (albeit convenient)
selection bias. This point will be discussed further. Some examples of
simulated cutouts are shown in Figure~\ref{fig:puppies}.

\section{Data mining}
\label{sect:datamining}
\begin{table} 
\centering
\begin{tabular}{ll}
\hline
symbol & meaning\\
\hline
$p$ & number of features per object \\
$K$ & number of classes of objects \\
$\mathbf{f}$ & feature vector (in $\mathbb{R}^{p}$) \\
$\mathbf{y}$ & membership probability vector (in $\mathbb{R}^{K}$) \\
$N_{t}$ & objects in a training set \\
$N_{v}$ & objects in a validating set \\
$R_{err}$ & error loss function \\
$R_{dev}$ & deviance loss function \\
$R_{reg}$ & regularization \\
$M$ & number of hidden nodes (in ANNs)\\
$\lambda$ & regularization parameter (in ANNs)\\
\hline
\end{tabular}
\caption{Nomenclature used here for our machine-learning techniques
(Section \ref{sect:techn1}).}
\label{tab:names}
\end{table}

In a {supervised clasification} problem, one is given a training
set of $N_t$ objects, each of which has a {probability vector}
whose entries are the probabilities of belonging to certain classes.
The aim is then to find a best fit to the probability vectors in the
training set, together with predictive power on other, new objects.
Many techniques have been developed for machine learning and
classification \citep[see][for a general review]{bal10}.
 We will briefly introduce two of them, Artificial Neural Networks (ANNs)
 and Gradient-Boosted Trees (GBTs), whose choice reflects just our personal preference (see also Section~\ref{sect:others}). 

In the lens detection problem we train these {learners} on simulated
lensed quasars. Their theoretical performance is evaluated on error and deviance metrics, described in Section~\ref{sect:errdev}.
On the other hand, the searches for lensed quasars are often biased to
bright objects with QSO-like photometry, which are not a complete
representation of the whole population (c.f. Figure~\ref{fig:colcolplots}). 
Then, we also evaluate the performance of the learners by testing them against a
sample of objects found in past searches, such as the SQLS
candidates. In particular, we will test our alogithms on the SQLS
morphologically selected sample of lensed QSO targets {from SDSS DR7
by \citet{ina12}.}

Regardless of the technique, however, a preliminary step of
{dimensional reduction} on the data is necessary. If we examine
10-arcsecond wide cutouts in $griz$ bands, then each source has a $25
\times 25$ image for $g, r, i,$ and $z$, implying a feature space of
2500 dimensions for the raw pixel values. This {curse of
dimensionality} presents a computational challenge, while also leading
to an increase in variance and degraded classifier performance.
Fortunately, there is significant structure in the images, so that the
information  can be compressed onto a lower-dimensional manifold.
Instead of using the raw data themselves, we first identify
{features}. Some of them are available at catalogue level and
reflect some rough physical intuition about the photometry and
morphology of the systems, others are {data driven}, i.e. can
be extracted by suitable combinations of pixel values from the cutouts
without imposing any physical intuition. This simply generalizes the
common procedure of drawing cuts and wedges in colour-magnitude space,
in which case the feature array $\mathbf{f}\in\mathbb{R}^{p}$ would
simply contain the magnitudes in different bands. Given our mining
strategy, we first apply techniques with features extracted from the
survey catalogue, then we use data-driven features and pattern
recognition on the cutouts of those systems that pass the first
selection stage. 

Different populations of objects can be separated by setting
boundaries in feature space\footnote{For the reader's convenience,
Table \ref{tab:names} summarizes the nomenclature introduced in this
Section.}. The accuracy of the classification depends on how many
boundaries are set and how flexible they are. The simplest separation
relies on linear and affine boundaries, i.e. the partition of feature
space ($\mathbf{f}\in\mathbb{R}^{p}$) in regions delimited by
hyperplanes, each with {weight vector} $\boldsymbol{\alpha}$
and {bias} $a_{0}:$
\begin{equation}
\boldsymbol{\alpha}\cdot\mathbf{f}+a_{0}=0\ .
\end{equation}
If different populations overlap in feature space, drawing many
boundaries enables the construction of membership probabilities.
Finally, a concatenation of classification steps enables the
construction of non-linear classifiers. The details of these
procedures are specific to different machine learning algorithms.

\subsection{Dimensional reduction: catalogue parameters}

\begin{figure*}
 \centering
 \includegraphics[width=0.33\textwidth]{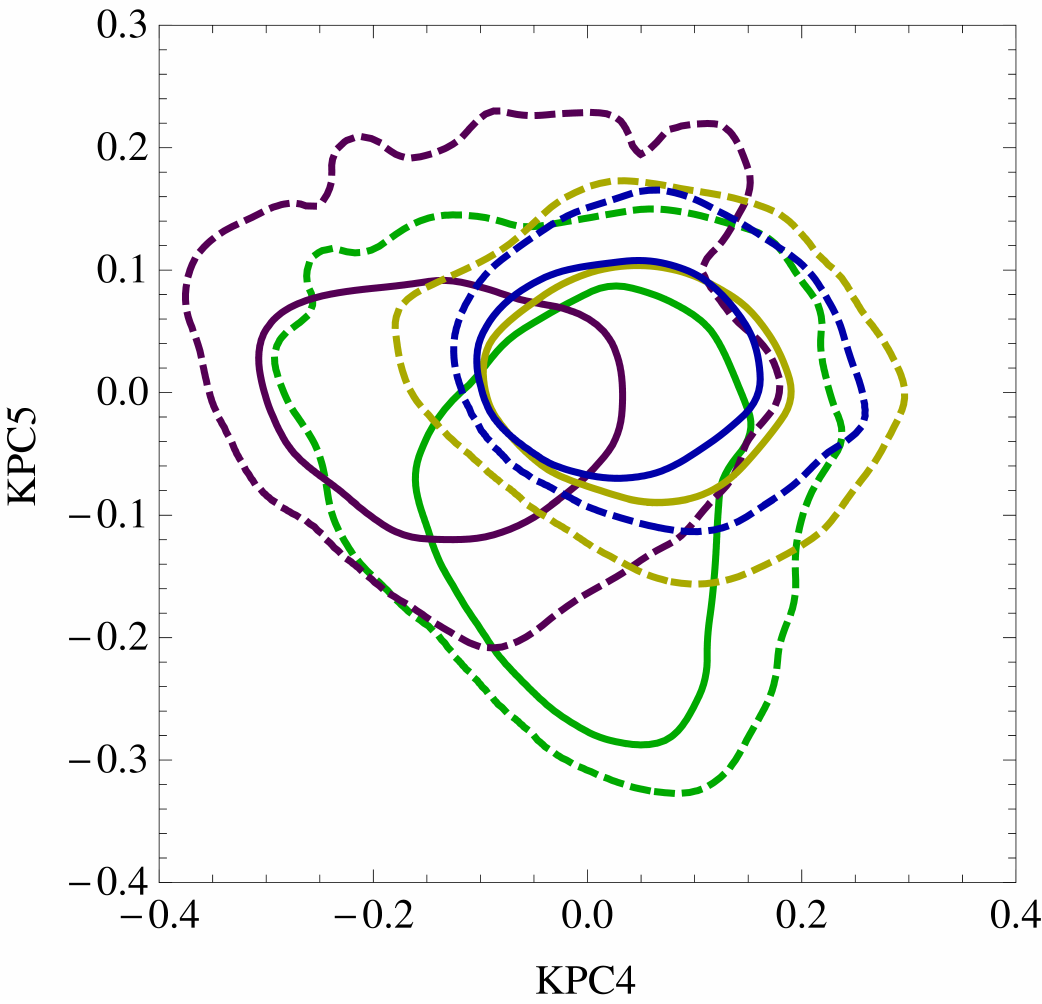}
 \includegraphics[width=0.33\textwidth]{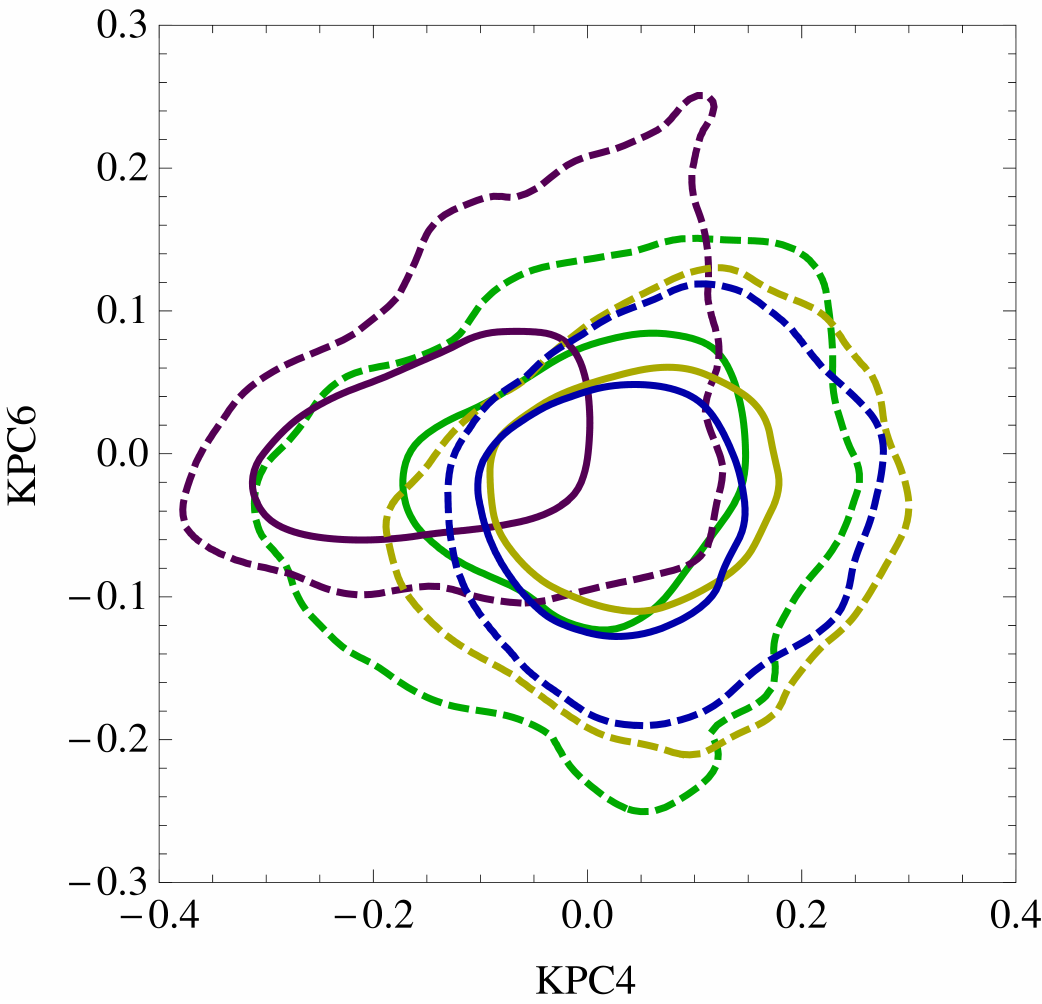}
 \includegraphics[width=0.33\textwidth]{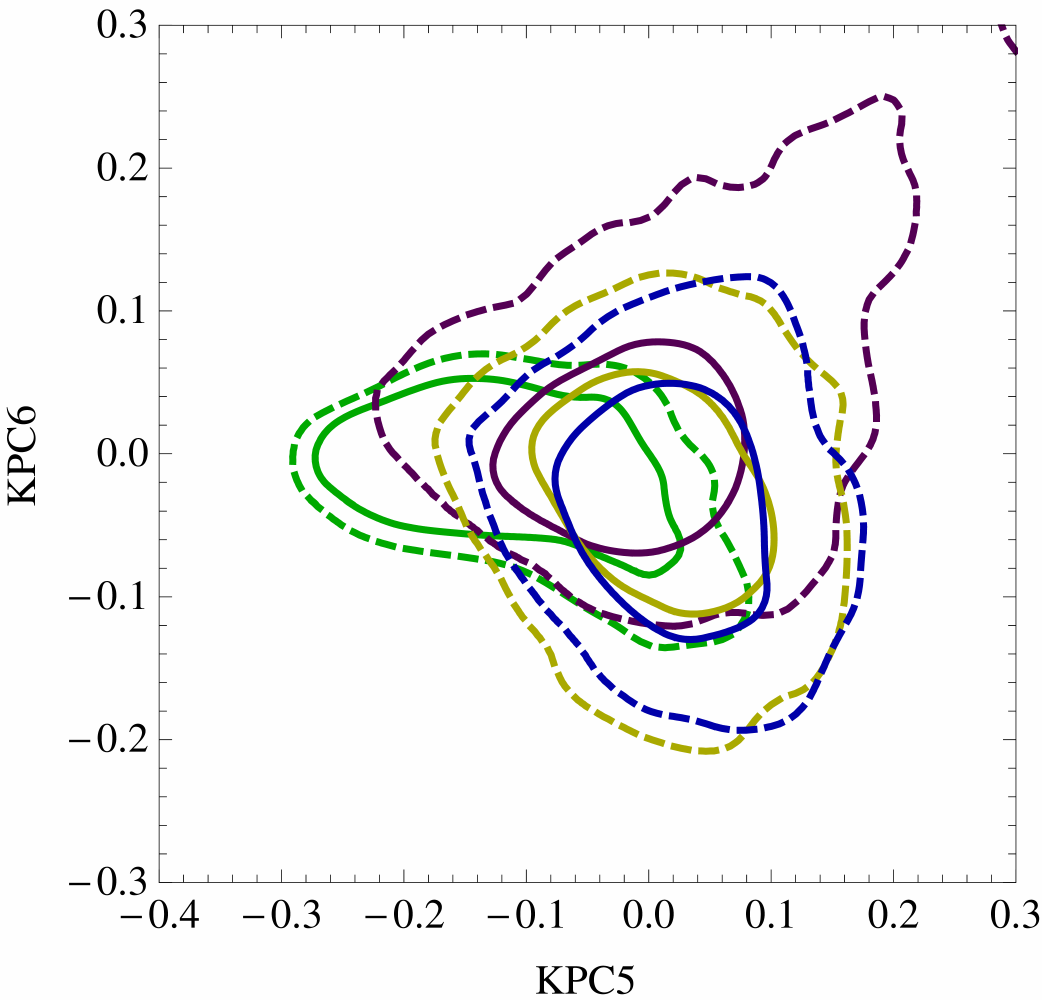}
\caption{\small{Correlations among the projections of the simulated
cutouts onto some of the first kernel-PCs. Contours delimit the
various simulated populations: lensed QSO (green), QSO+LRG (purple),
2QSO (yellow) and single QSOs (blue). The label $\mathrm{KPC}j$
denotes the $j-$th kernel principal component. The populations extend
in different directions and partially overlap in kernel-PC space. The
separation into classes is attained by modelling the first 200
kernel-PCs simultaneously.}}
\label{f-kpca}
\end{figure*}

When skimming a whole catalogue for targets, we generalize the SQLS
idea of searching for objects with promising photometry and
morphological information.  In this work, we exploit the magnitudes in
$griz$ bands (AB system) and WISE $W1,W2$ bands (Vega system) for the
photometric information. The morphology is encoded via the second
moments, and from these, the axis ratios and position angles (p.a.s)
in the four $griz$ bands. The underlying idea is that the quasar
images and the lensing galaxy, when blended together, will still have
some distinctive features in the morphology and in how it varies with
observing band. For example, in the case of a double we may expect the
red deflector to be in the middle and be more relevant in the redder
bands. Therefore the elongation should decrease with wavelength,
barring uncertainties from noise and pixel-size. Since one p.a. is
arbitrary, we use the three differences
$(\delta{\phi}_{r},\delta{\phi}_{i},\delta{\phi}_{z})$ between the
p.a. in $g$ band and those in the other ones. In the end, this leaves
us with $p=13$ features per object, thus mapping the
highly dimensional space of raw pixels onto a 13-dimensional feature
space.

\subsection{Dimensionality reduction: Kernel PCA}
\label{s-dim_reduce}
Once the targets have been identified in catalog space, we can return to the image pixels
 and extract more features beyond the simple photo-morphological information used in the previous step.
 A number of techniques can be deployed to extract data-driven features from the raw pixels in the image cutouts.

We investigated both principal component analysis (PCA) and kernel
principal component analysis \citep[KPCA,][]{Scholkopf1998a} on 25 by
25 pixel cutout images.  
In PCA, the feature vectors are expressed as
combinations of the eigenvectors of the data covariance matrix. This
is useful in order to isolate the directions of maximum variance and
set simple boundaries in feature space. KPCA is similar to PCA, but it
uses a kernel $k$ and a map $\Phi$ to embed the feature space
$\mathbb{R}^{p}$ into a higher-dimensional space $\mathcal{H},$ such
that
\begin{equation}
 k(\mathbf{x},\mathbf{y})=\langle\Phi(\mathbf{x}),\Phi(\mathbf{y})\rangle
\end{equation} 
is a scalar product in $\mathcal{H},$ and then perform the PCA there.
There is no need to compute the map
$\Phi:\mathbb{R}^{p}\rightarrow\mathbb{R}^{D}$ explicitly -- a fact
known in jargon as {kernel trick}. Because of this, KPCA can
perform non-linear dimension reduction, while PCA is a linear
dimension reduction technique. Further details on KPCA are given in
the Appendix and can be found in \citet{Scholkopf1998a}. 

For PCA, we found that the first 500 principal components contain
about $90\%$ of the variance in the raw pixel feature space.
% The first ten PCs of the simulated sample are shown in Figure \ref{f-pca}.
We can further reduce the number of components by means of the kernel
trick.
% Different functional forms can be used for the kernel in KPCA.
To this aim we used the {radial basis function} kernel,
\begin{equation}
k(\mathbf{x},\mathbf{y})\ \propto\ \exp\left[-||\mathbf{x}-\mathbf{y}||^{2}/(2\delta^{2})\right]
\end{equation}
%  which is equivalent to a Gaussian function in the Euclidean distance between data points
in the 2500-dimensional feature space. 
% PJM: should this be "2500-dimensional pixel space"?
The width $\delta$ of the kernel is a tuning parameter. We
experimented with a few different values of the kernel width, and
found that a value of 0.25 times the median nearest-neighbour distance
gave good separation of the classes, as projected onto the first
several KPCs. We did not tune this parameter further, choosing rather
to use our degrees of freedom for tuning the classifiers as described
below. For reference, the projections of the data set classes onto
some KPCs are shown in Figure~\ref{f-kpca}.
\subsection{Error and Deviance Metrics}
\label{sect:errdev}
In supervised classification, the performance of an algorithm can be quantified by measuring how its output
 probability vectors $\mathbf{t}_{i}$ deviate from the true ones $\mathbf{y}_{i}$ ($i=1,...,N$ running over a sample of $N$
 objects). This can be simply estimated via the error loss function,
\begin{equation}
R_{err}=\frac{1}{N}\sum\limits_{i=1}^{N}||\mathbf{y}_{i}-\mathbf{t}_{i}||^{2}\ .
\label{eq:Rerr}
\end{equation}
Another commonly used loss function is the deviance, defined as the
conditional entropy of the output probabilities with respect to the
true ones:
\begin{equation}
R_{dev}\ =-\sum\limits_{i=1}^{N}\sum\limits_{k=1}^{K}y_{i,k}\log(t_{i,k})\ 
\label{eq:Rdev}
\end{equation}
($K$ being the number of distinct classes).
The classifier algorithms are {trained} to minimize either
 $R_{err}$ or $R_{dev}$ over suitable training and validating sets,  as described in Sect.~\ref{sect:techn1},~\ref{sect:techn2}
  below. The error and deviance metrics will also be used to assess the reliability of the learners, once they are trained.
  
\subsection{Techniques: Artificial Neural Networks}
\label{sect:techn1}
\begin{table} 
\centering
\begin{tabular}{|c|c|c|}
\hline
class & training & validating\\
\hline
l.QSO & 245 & 100\\
2QSO & 150 & 80\\
QSO+LRG & 265 & 100\\
BC & 90 & 75\\
\hline
\end{tabular}
\caption{Objects in the training and validation sets for each class.
When training the ANNs, the error and deviance metrics are evaluated
on ten different validating sets, so as to have a better grasp on
sample-to-sample variance.}
\label{tab:testval}
\end{table}
For target selection, we use Artificial Neural Networks (ANN). In the
simplest ANN scheme, given the feature vectors $\mathbf{x}_{i}$ in
the test set ($i=1,...,N_t$), the probability vectors
$\mathbf{y}_{i}\in\mathbb{R}^{k}$ are fit by combinations
\begin{equation}
\mathbf{t}_{i}=\sum\limits_{m=1}^{M}\boldsymbol{\beta}_{m}g(\boldsymbol{\alpha}_{m}\cdot\mathbf{x}_{i}+a_{0,m})
\label{eq:elmapp}
\end{equation}
with every weight  $\boldsymbol{\beta}_{m}\in\mathbb{R}^{K}$ and
$\boldsymbol{\alpha}_{m}\in\mathbb{R}^{p},$ over a {layer} of
$M$ {hidden nodes}, where $g$ is a smooth {activation
function} such that $g(\pm\infty)=1/2\pm1/2.$ One further passage is
made in order to ensure that the entries of each $\mathbf{t}_{i},$
being membership probabilities, be positive and sum to unity. Appendix
B gives a detailed description of the ANN architecture. The ANN is
 trained to minimize the error metric (eq.~\ref{eq:Rerr}).
The deviance metric (\ref{eq:Rdev}) is considered in a subsequent stage.

We also create ten {validating} sets, where $R_{err}$ is
computed but not minimized. We use {early stopping}, i.e.
interrupt the training when the mean error on the validating sets
stops decreasing. This is commonly interpreted as a symptom that the
learner is becoming {greedy} to imitate the training set,
whilst not improving in predicting the classification of objects in
the validating set.  The number of objects per class in training and
validating sets is given in Table \ref{tab:testval}. The different
proportions of objects were adapted so that the machines would learn
to correctly recognise most of them, especially the lensed quasars.
For this reason, the fraction of lensed quasars is slightly higher
than in the SQLS target sample.

To avoid overfitting, a regularization term
\begin{equation}
R_{reg}=\lambda\sum_{m}\left(||\boldsymbol{\alpha}_{m}||^{2}+||\boldsymbol{\beta}_{m}||^{2}+a_{0,m}^{2}\right)
\end{equation}
is added to $R_{err}.$ The performance of the ANN will ultimately
depend on $M$ and $\lambda$ (see Sect.\ref{sect:ANNtarg}).

A faster, albeit less accurate, alternative to ANNs are Extreme Learning Machines
 \citep[ELMs,][]{elms}, where eq.~(\ref{eq:elmapp})
 is used directly and just the $\boldsymbol{\beta}$ weights are optimized. 
The main advantage of ANNs over ELMs is that the latter's output are
not necessarily probability vectors, i.e. they do not always have
positive entries summing to unity. This can be troublesome when new
objects are considered, since the ELMs could extrapolate the outputs
to values well outside the $[0,1]$ range. The amenable property of
ANNs to output probability vectors is the main reason why we have
preferred them over ELMs for target selection.

\subsection{Techniques: Gradient-Boosted Trees}
\label{sect:techn2}

Gradient boosting \citep{Friedman2001a} is a general machine learning
technique that produces a prediction model using an ensemble of weak
learners, where a weak learner is a simple predictive model that may
only do slightly better than random guessing. The gradient boosting
classifier is built up slowly by fitting a sequence of weak learners
to minimize a loss function, which for classification is typically
chosen to be the deviance (\ref{eq:Rdev}). The output for the
predictive model is the ensemble average of the prediction for each
weak learner. At each iteration of the algorithm, a new weak learner
is trained on the residuals for the current model, and this weak
learner is added to the ensemble with a contribution proportional to a
learning rate parameter. The tuning parameters for the gradient
boosting algorithm are the learning rate and the number of weak
learners in the ensemble. When the learning rate is smaller, the model
is built up more slowly and a larger number of weak learners is
needed.  Smaller learning rates tend to lead to better test error as
they build up the model in a more controlled manner, although they
lead to longer computations as they require a higher number of
learners. Gradient boosting has been found to be powerful and robust
in a variety of different prediction problems
\citep[e.g.,][]{Hastie2009a}, and is very slow to overfit. Further
details are given in the Appendix, as well as \citet{Friedman2001a}
and \citet{Hastie2009a}.

In our case we use shallow decision trees for the weak learners. A
decision tree is made up of a set of binary splits that partition the
feature space into a set of constant functions. For classification,
the output from the tree is the probability that a data point belongs
to a given class given the partition of the input feature space that
the data point falls into. The aim is to approximate membership probabilities by
 piecewise-constant functions in regions $R_{j}$ of feature space,
\begin{equation}
t(\mathbf{x})=\sum\limits_{m=1}^{M}\nu\sum\limits_{j=1}^{J}\gamma_{j,m}I(\mathbf{x}\in R_{j})
\end{equation}
(see Appendix~\ref{sect:appGBTs}), over $M$ trees of depth $J.$

% For example, the first split in the tree
%could be, say, that all sources with $g < 20$ go into one partition,
%while all sources with $g \geq 20$ are placed in a different
%partition. Then, for the $g < 20$ partition we could make another
%partition at $r = 19$, such that if $r < 19$ there is a $90\%$
%probability that a source is a lens, while for $r \geq 19$ there is
%only a $10\%$ probability that the source is a lens. A similar split
%would occur for the $g \geq 20$ partition at a different value of $r$,
%or for an entirely different feature. Moreover, we need not stop at
%this point, but could partition the space even more by growing the
%tree to a larger depth.

 Within the context of gradient boosting, the
class probabilities are obtained from the average over the ensemble of
shallow decision trees via the learning rate $\nu.$ Because of this, the depth of the trees is an
additional tuning parameter in this classification model.

In addition, in our analysis we use a stochastic variant of the
original gradient boosting algorithm \citep{Friedman2002a}. In
stochastic gradient boosting only a random subsample of the training
data is used at each iteration to build the decision tree; note that
this subsample is randomly chosen at each iteration of the gradient
boosting algorithm, and is not constant throughout the algorithm.
\citep{Friedman2002a} found empirically that random subsampling tended
to improve the test errors. In addition, we use random subsampling
because it enables us to monitor the deviance loss on the subset of
the data that is not used in the fit at each iteration. This yields an
estimate of the prediction error of the model as a function of the
number of trees, and we choose the number of trees to minimize this
estimated prediction error. Other tuning parameters are typically
chosen using cross-validation.

\subsection{Techniques: Cross-validation}
\label{s-cv}

Cross-validation is a statistical technique for estimating the
prediction error of a model. The basic idea is to divide the training
set into $K$ separate subsamples. Then, one subsample is withheld and
the model is trained on the remaining $K-1$ subsamples. The prediction
error from this model is calculated for the subsample that was
withheld from training the model. The procedure is repeated for each
of the other $K-1$ subsamples as well, and the cross-validation error
is obtained as the average prediction error over the subsamples.

%While cross-validation may be used to obtain an estimate of the test error, \textbf{it is typically a biased estimate
% [AA: ?!]}.
For most applications, the tuning parameters are chosen to minimize
the cross-validation error. In addition, cross-validation is often
used to choose the number of features to include in a regression or
classification model. Finally, it also helps control overfitting, by
attempting to find the tuning parameters and feature set that minimize
an average out-of-sample prediction error.

\begin{figure*}
 \centering
 \includegraphics[width=0.1\textwidth]{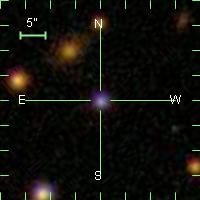}
 \includegraphics[width=0.1\textwidth]{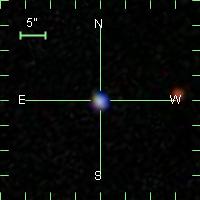}
 \includegraphics[width=0.1\textwidth]{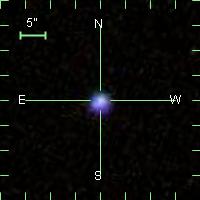}
 \includegraphics[width=0.1\textwidth]{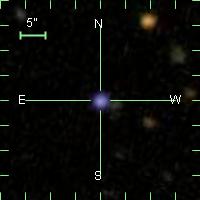}
 \includegraphics[width=0.1\textwidth]{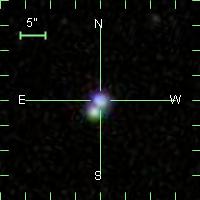}
 \includegraphics[width=0.1\textwidth]{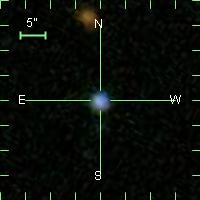}
 \includegraphics[width=0.1\textwidth]{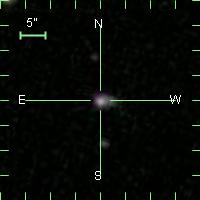}
 \includegraphics[width=0.1\textwidth]{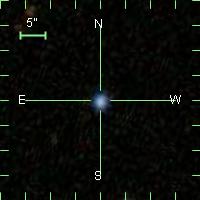}
 \includegraphics[width=0.1\textwidth]{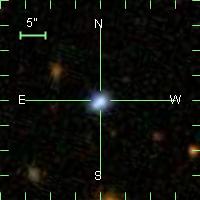}\\
 \includegraphics[width=0.1\textwidth]{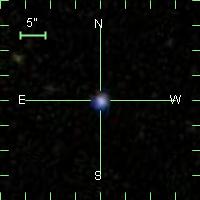}
 \includegraphics[width=0.1\textwidth]{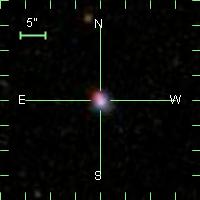}
 \includegraphics[width=0.1\textwidth]{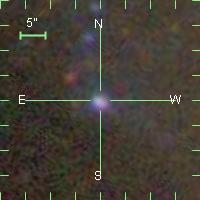}
 \includegraphics[width=0.1\textwidth]{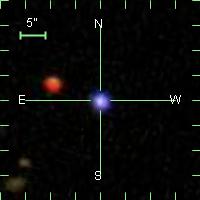}
 \includegraphics[width=0.1\textwidth]{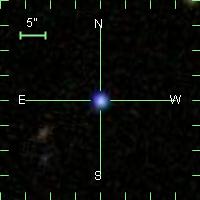}
 \includegraphics[width=0.1\textwidth]{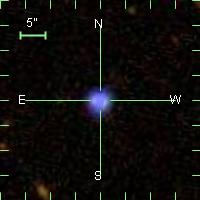}
 \includegraphics[width=0.1\textwidth]{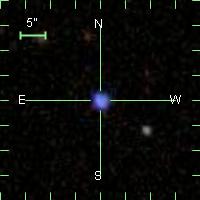}
 \includegraphics[width=0.1\textwidth]{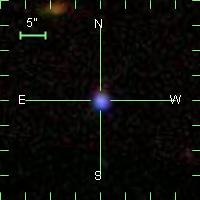}
 \includegraphics[width=0.1\textwidth]{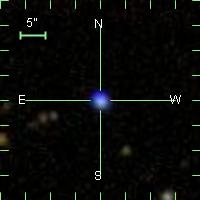}\\
 \includegraphics[width=0.1\textwidth]{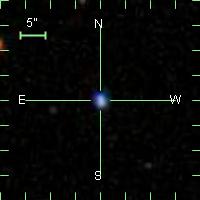}
 \includegraphics[width=0.1\textwidth]{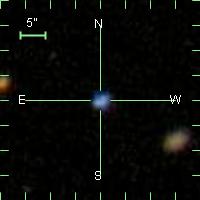}
 \includegraphics[width=0.1\textwidth]{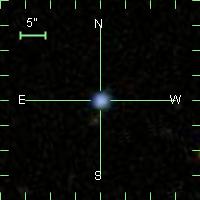}
 \includegraphics[width=0.1\textwidth]{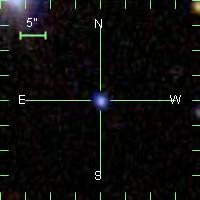}
 \includegraphics[width=0.1\textwidth]{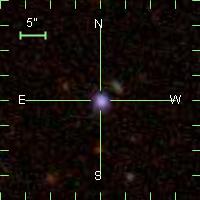}
 \includegraphics[width=0.1\textwidth]{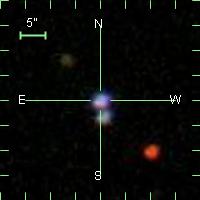}
 \includegraphics[width=0.1\textwidth]{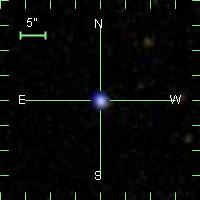}
 \includegraphics[width=0.1\textwidth]{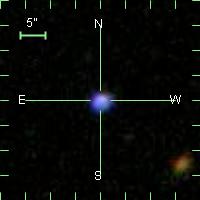}
 \includegraphics[width=0.1\textwidth]{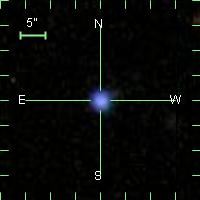}\\
 \includegraphics[width=0.1\textwidth]{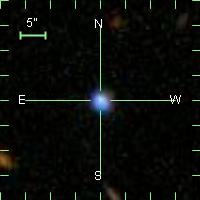}
 \includegraphics[width=0.1\textwidth]{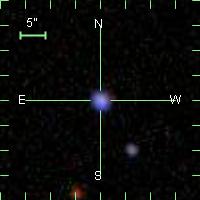}
 \includegraphics[width=0.1\textwidth]{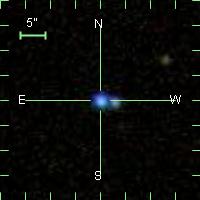}
 \includegraphics[width=0.1\textwidth]{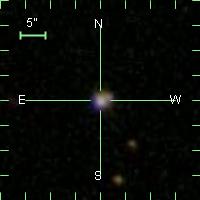}
 \includegraphics[width=0.1\textwidth]{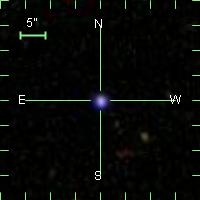}
 \includegraphics[width=0.1\textwidth]{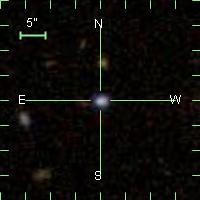}
 \includegraphics[width=0.1\textwidth]{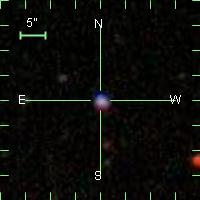}
 \includegraphics[width=0.1\textwidth]{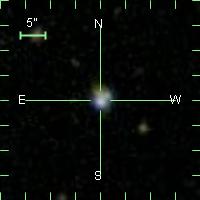}
 \includegraphics[width=0.1\textwidth]{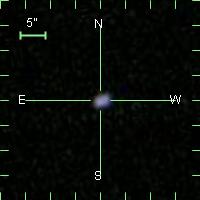}\\
 \includegraphics[width=0.1\textwidth]{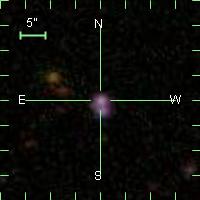}
 \includegraphics[width=0.1\textwidth]{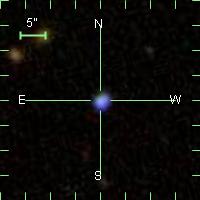}
 \includegraphics[width=0.1\textwidth]{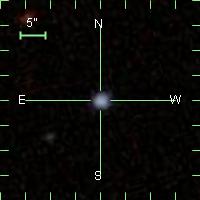}
 \includegraphics[width=0.1\textwidth]{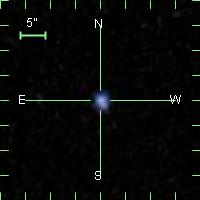}
 \includegraphics[width=0.1\textwidth]{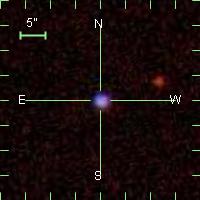}
 \includegraphics[width=0.1\textwidth]{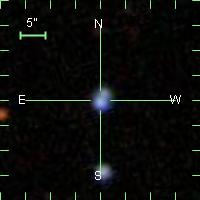}
 \includegraphics[width=0.1\textwidth]{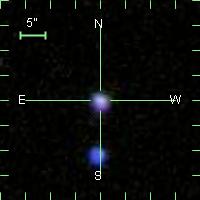}
 \includegraphics[width=0.1\textwidth]{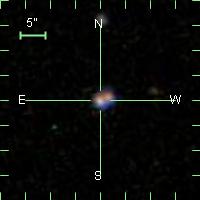}
 \includegraphics[width=0.1\textwidth]{pics/SQLSpics/45.jpeg}\\
 \includegraphics[width=0.1\textwidth]{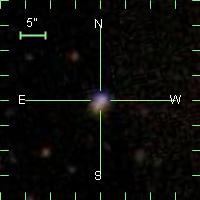}
 \includegraphics[width=0.1\textwidth]{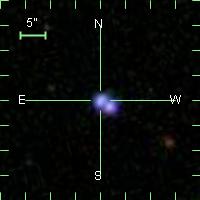}
 \includegraphics[width=0.1\textwidth]{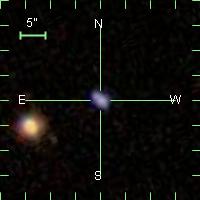}
 \includegraphics[width=0.1\textwidth]{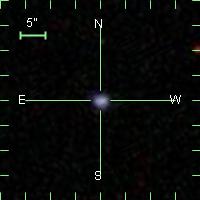}
 \includegraphics[width=0.1\textwidth]{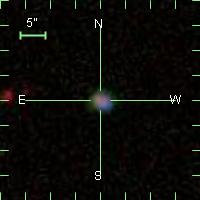}
 \includegraphics[width=0.1\textwidth]{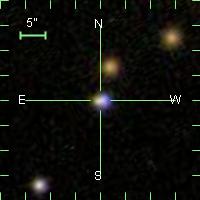}
 \includegraphics[width=0.1\textwidth]{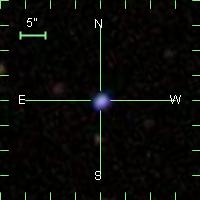}
 \includegraphics[width=0.1\textwidth]{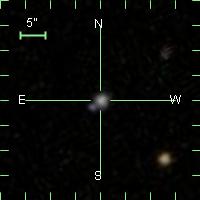}
 \includegraphics[width=0.1\textwidth]{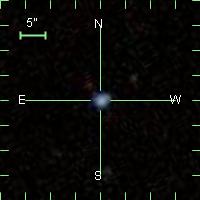}
\caption{\small{The morphologically-selected candidates in SDSS DR7 by
\citet{ina12}.}}
\label{fig:SQLSmontage}
\end{figure*}

\section{Results}
\label{sect:results}

{For each of the two steps, target selection and candidate selection,
we trained and validated the appropriate learning algorithm from 
Section~\ref{sect:datamining} using the relevant dataset as defined in
Section~\ref{sect:data}, optimizing their structural parameters by
cross-validation. We now quantify the} performance of each learner in terms of
classification error statistics, deviance, false-vs-true positive
rates and purity-vs-completeness, as specified below. Our metrics are
built both on the training/validation data, 
and on the SQLS sample of lensed quasars and false positives 
\citep{ina12}. While examples of the former are displayed in
Figure~\ref{fig:puppies}, here we show the SQLS objects in
Figure~\ref{fig:SQLSmontage}. Once again, the blended nature and
similar colours of the objects are the main obstacles to an efficient and
simple separation of true and false positives.

\begin{table} 
 \centering
\begin{tabular}{|c|c|c|c|c|c|}
\hline
$\lambda$ & $M$ & $error$ & $deviance$ & \multicolumn{2}{c}{SQLS first test} \\
 &  & per system & per system & purity & completeness \\
\hline
$\lambda=0$  & 13 & $0.56\pm0.03$ & $0.63\pm0.07$ & 0.5 & 0.625 \\
 & 17 & $0.59\pm0.03$ & $0.74\pm0.09$ & 0.22 & 0.225 \\
 & 20 & $0.62\pm0.04$ & $0.69\pm0.11$ & 0.30 & 0.5 \\
 & 23 & $0.56\pm0.04$ & $0.64\pm0.14$ & 0.5 & 0.625 \\
 & 25 & $0.57\pm0.03$ & $0.61\pm0.09$ & 0.45 & 0.625 \\
\hline
$\lambda=0.01$  & 13 & $0.61\pm0.03$ & $1.10\pm0.44$ & 0.33 & 0.25 \\
 & 17 & $0.58\pm0.04$ & $0.84\pm0.36$ & 0.44 & 0.5 \\ 
 & 20 & $0.58\pm0.04$ & $0.94\pm0.42$ & 0.475 & 0.375 \\ 
 & 23 & $0.57\pm0.04$ & $0.96\pm0.42$ & 0.42 & 0.375 \\ 
 & 25 & $0.60\pm0.03$ & $0.76\pm0.10$ & 0.42 & 0.62 \\ 
\hline
$\lambda=0.2$  & 13 & $0.56\pm0.03$ & $0.64\pm0.15$ & 0.375 & 0.75 \\
  & 17 & $0.56\pm0.03$ & $0.65\pm0.15$ & 0.46 & 0.75 \\
  & 20 & $0.56\pm0.03$ & $0.64\pm0.15$ & 0.4 & 0.625 \\
  & 23 & $0.56\pm0.03$ & $0.62\pm0.13$ & 0.4 & 0.75 \\
  & 25 & $0.56\pm0.03$ & $0.69\pm0.20$ & 0.35 & 0.625 \\
\hline
$\lambda=0.5$  & 13 & $0.57\pm0.03$ & $0.59\pm0.05$ & 0.3 & 1 \\
 & 17 & $0.57\pm0.03$ & $0.59\pm0.05$ & 0.3 & 0.875 \\
 & 20 & $0.57\pm0.03$ & $0.59\pm0.05$ & 0.4 & 0.875 \\
 & 23 & $0.58\pm0.02$ & $0.62\pm0.04$ & 0.4 & 0.875 \\
 & 25 & $0.57\pm0.02$ & $0.59\pm0.05$ & 0.35 & 0.75 \\
\hline
$\lambda=1$  & 13 & $0.59\pm0.02$ & $0.65\pm0.04$ & 0.3 & 1 \\
 & 17 & $0.59\pm0.02$ & $0.65\pm0.04$ & 0.3 & 1 \\
 & 20 & $0.59\pm0.02$ & $0.65\pm0.04$ & 0.25 & 1 \\
 & 23 & $0.59\pm0.02$ & $0.65\pm0.04$ & 0.30 & 1 \\
 & 25 & $0.59\pm0.02$ & $0.65\pm0.04$ & 0.25 & 1 \\
 \hline
\end{tabular}
\caption{Performance of the ANNs with different choices of the
regularization parameter $\lambda$ and number of nodes $M,$ according
to different quantifiers. Columns `error' and `deviance'  list the
mean loss quantifiers $\sqrt{R_{err}}$ and $R_{dev}$ over the
validation set, run on ten different validation sets. The `SQLS'
columns list the purity and completeness for the sample of SQLS
objects whose output probabilities satisfy $p(QSO+LRG)<0.35,$
$p(BC)<0.35,$ $p(2QSO)<0.8,$ without any further restriction on
$p(l.QSO).$ }
\label{tab:ANN}
\end{table}

\begin{figure}
 \centering
 \includegraphics[width=0.45\textwidth]{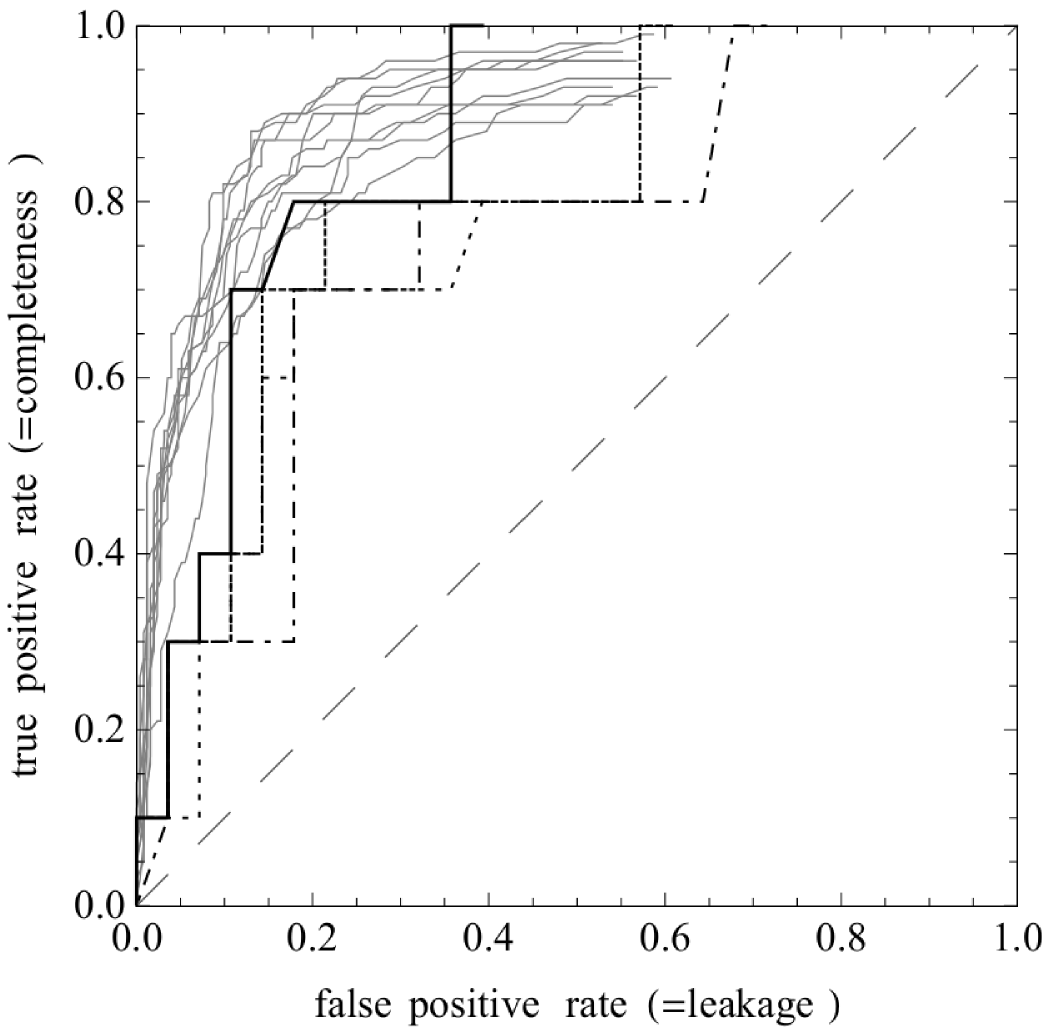}\\
 \includegraphics[width=0.45\textwidth]{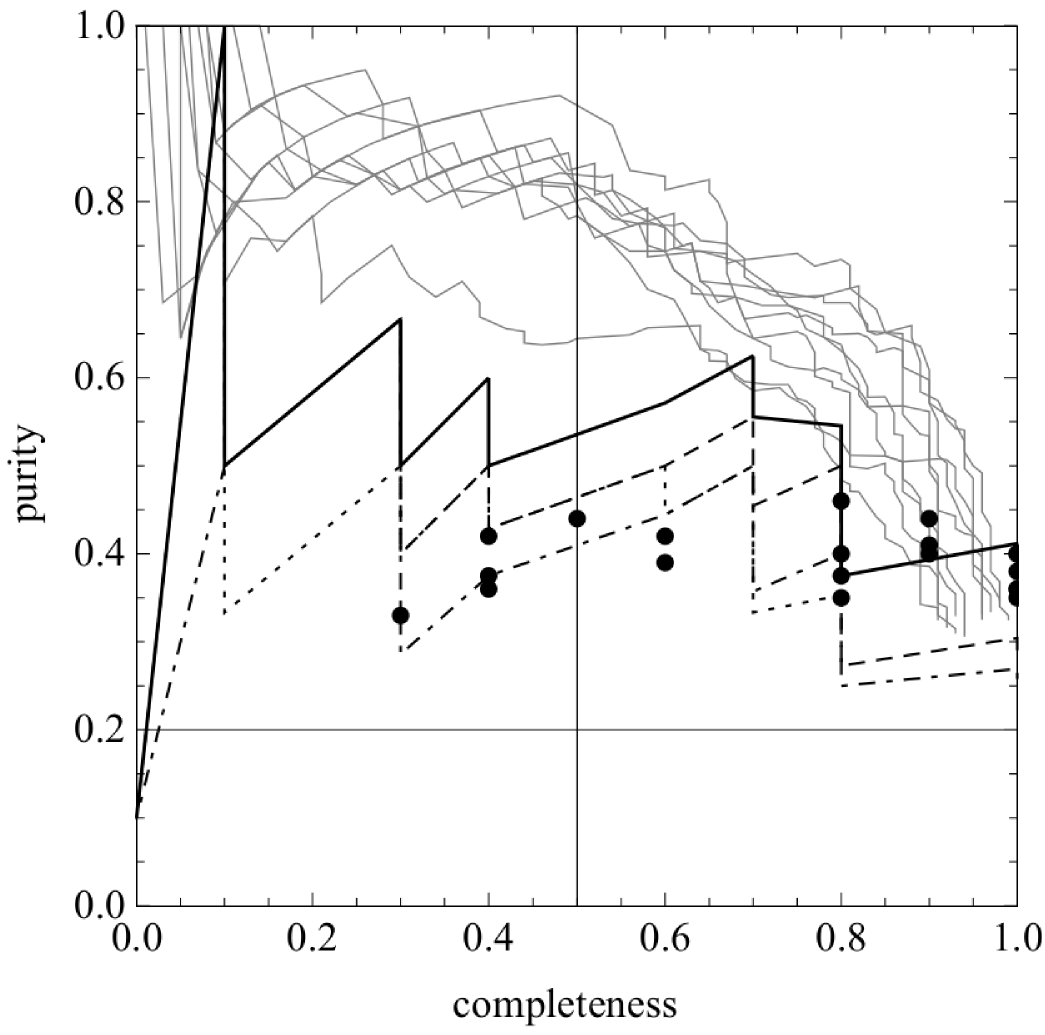}
\caption{\small{Performance of the target selection, with progressive
cuts in the output probabilities, displayed as a receiver operating
characteristic curve (top) or purity-completeness plot (bottom). The
long-dashed, 1:1 straight line in the bottom panel marks the
performance of random classifiers. Grey lines: performance of an ANN
with $M=13$ and $\lambda=1$ on the ten validating sets. Other lines
mark the performance of the four best ANNs: full (resp short-dashed,
dotted, dot-dashed) stands for $M=13$ and $\lambda=1$ (resp. $M=13$
and $\lambda=0.5,$ $M=20$ and $\lambda=0.5,$ $M=20$ and $\lambda=1$).
Bullets mark the performance of other ANNs, accepting all objects
regardless of $p(l.QSO).$ }}
\label{fig:ANNperf}
\end{figure}

\subsection{Target Selection with Neural Networks}
\label{sect:ANNtarg}

The structural parameters of the ANN we used for lens target selection
are the number $M$ of nodes and the regularisation parameter
$\lambda.$ We varied $\lambda$ between $0$ and $1,$ whilst $M$ was
varied between $13$ and $25.$ The results are shown in
Table~\ref{tab:ANN}, where the performance is quantified by means of
four metrics explained below. These are: rms error and deviance on the
validating sets, and purity and completeness when run on the SQLS 
{morphologically selected objects \citep{ina12}.

The performance on the validating set is quantified by the
classification error and deviance per system. The error per system is
defined as the r.m.s. distance of the output probability vectors from
their true value, for points in the validating set, i.e.
$\sqrt{R_{err}}$ (from eq.~\ref{eq:Rerr}) for each validating set.
Similarly, the deviance per system is $R_{dev}/N_{v}$ (from eq.~\ref{eq:Rdev}).

When training the ANNs, we followed the behaviour of the error and the
deviance on ten different validating sets. This allowed us to assess
how the performance of a classifier would vary in practice, if a
different validating set were chosen. The values with errorbars in
Table~\ref{tab:ANN} then show the average performance on a validating
set and its typical (r.m.s.) variability.

The performance on the SQLS objects is measured in terms of the purity
and completeness of the selected targets. Alternatively the true- and
false-positive rates can be considered, being the fraction of true or
false positives that are flagged as possible lensed quasar (l.QSO)
targets by the ANNs. For those, one further step is necessary. Since
the classifier outputs probabilities for each object to belong to any
of the classes, one needs to define a probability threshold in order
to select a system as a target or reject it. Objects with
$p(QSO+LRG)>0.35,$ $p(BC)>0.35,$ $p(2QSO)>0.8,$ are always flagged as
non-targets and rejected.  
Such a skimming produces a sample of putative lensed QSOs, whose
purity and completeness can be used to quantify the efficiency of the
ANNs to separate into classes when presented with real data. These are
listed in the last column of  Table~\ref{tab:ANN}.

Finally, a minimum threshold must be chosen in order to select just
those candidates whose output $p(l.QSO)$ is sufficiently high. By
varying the acceptance probability threshold, we can increase the
purity of the target set at the expense of completeness -- or vice
versa. Figure \ref{fig:ANNperf} shows the {receiver operating
characteristic} (ROC), i.e. the relation between true positive rate
and false positive rate as the acceptance threshold on $p(l.QSO)$ is
varied. This figure also shows a plot of purity versus completeness.
The black lines display the performance of the best four ANNs (cf
Table \ref{tab:ANN}) on the SQLS test objects. The bullets mark the
performance of all the other ANNs at maximum completeness, i.e.
without cuts in $p(l.QSO).$ For illustrative purposes, the ten grey lines in each panel show the
performance of  an ANN with $M=13$ and $\lambda=0.5$ on the ten
validating sets. When plotting these, the weight of lensed quasars in
the validating set has been slightly reduced, in order to match the
$20\%$ overall fraction of true positives in SQLS and offer a fair
comparison.

\subsection{Gradient-Boosted Trees and candidate selection}

We trained a classifier directly on the pixel-by-pixel values of the images using several different algorithms.
 The first step in this process was to normalize all of the images such that the sum of their flux values for each pixel
 over all of the observational bands was equal to unity. This normalization implies that the classifier will focus on the
 colour-morphological information for each source. Then, we randomly split the set of simulated images into a training
 and test set, where the test set contained $25\%$ of the sources. The test set was used to provide an estimate of
 the test error and was set aside until the end of the analysis; it was not used in the dimension reduction step, nor in
 the training step.
\begin{table}
\begin {center}
\footnotesize
\begin{tabular}{|r|cccc|}
\hline
\multicolumn{5}{c}{Unbiased simulated sample.} \\
\hline
\multicolumn{1}{c}{} & \multicolumn{4}{c}{percentage classified as...} \\
  & l.QSO & 2QSOs & QSO+ETG & s.QSO  \\
 true l.QSO & 88.0 & 6.7 & 4.9 & 0.4 \\
 true 2QSO & 6.1 & 57.5 & 7.3 & 29.1 \\
 true QSO+ETG & 6.6 & 5.8 & 84.7 & 2.9 \\
 true s.QSO & 2.4 & 13.8 & 3.2 & 80.6 \\
\hline
\hline
\multicolumn{5}{c}{Biased simulated sample (eq.s~\ref{eq:cuts}).} \\
\hline
\multicolumn{1}{c}{} & \multicolumn{4}{c}{percentage classified as...} \\
  & l.QSO & 2QSOs & QSO+ETG & s.QSO  \\
 true l.QSO & 84.5 & 6.5 & 8.6 & 0.4 \\
 true 2QSO & 2.7 & 84.2 & 10.9 & 2.2 \\
 true QSO+ETG & 6.1 & 15.3 & 77.1 & 1.5 \\
 true s.QSO & 2.6 & 59.5 & 10.3 & 27.6 \\
\hline
\end{tabular}
\end{center}
\vspace{-0.3cm}
\caption{ Confusion matrix for classification of the simulated cutouts with gradient-boosted trees, using KPCA for the
 dimensional reduction. Top sub-table: unbiased simulated samples; bottom: biased samples following the cuts in eq.s~(\ref{eq:cuts}).
}
\label{tab:confmat}
\end{table}
\begin{table}
\begin {center}
\footnotesize
\begin{tabular}{|r|cc|}
\hline
dim.red. & purity & completeness \\
\hline
 PCA (500) & 6/8 (=75\%) & 6/10 (=60\%) \\
 KPCA (200) & 7/10 & 7/10   (=70\%)\\
\hline
\end{tabular}
\end{center}
\vspace{-0.3cm}
\caption{ Performance of Gradient-Boosted Trees, trained on biased simulated samples, when classifying
 targets (from Sect.\ref{sect:techn1}) in the SQLS dataset.
}
\label{tab:finalGBTs}
\end{table}
\subsubsection{Training and Testing the Gradient-Boosted Decision Trees}
\label{s-gbc}
We used the stochastic gradient boosting algorithm, with decision tree classifiers as the base weak learners,
 to train a classifier using the first 200 KPCs derived from the pixel-level information in the images.
 We set the learning rate to be $\nu = 0.01$, a typical value. At each stage 
 we trained the new tree on $80\%$ of the data, while the remaining $20\%$ were used to monitor the validation error.
 The number of trees in the ensemble were chosen to minimize this validation error, leading to an ensemble of 1190 trees.
 Since the learning rate is intertwined with the number of trees in the ensemble, there is little to gain from tuning both.
 The depth of the trees was chosen to minimize the 7-fold cross-validation misclassification error rate,
 leading to trees with a depth of 5 nodes.
 
Having exploited much of the photo-morphological information when selecting targets,
 the classification bottleneck is encountered at this stage.
% We first ran the classifiers on cutouts from
% the same simulated dataset as in Sect.\ref{sect:ANNtarg} and they were effective in separating the four different classes,
% as summarised by the confusion matrix in Table\ref{tab:confmat}. However, the algorithm performed poorly on the
% SQLS dataset, classifying every object as a QSO pair or lensed QSO, thus missing the single quasars and QSO+LRG
% alignments.
 Also, there is a clear selection bias in the SQLS sample, towards objects whose photometry resembles
 closely that of bright quasars or quasar pairs, as is evident from fig.\ref{fig:colcolplots}. To account for that,
 we trained the classifiers on cutouts of simulated objects whose photometry satisfied
 the following, common cuts in colour-magnitude space\footnote{As usual,
$griz$ magnitudes in the AB system, WISE in the Vega system.}:
\begin{eqnarray}
\nonumber 16<i<20,\ g-r<0.6,\ r-i<0.45,\ i-z<0.4,\\
\nonumber 2.5<i-W1<5,\ 0.5<W1-W2<1.5,\\
g-i<1.2(i-W1)-2.8\ .
\label{eq:cuts}
\end{eqnarray}
Table \ref{tab:confmat} shows the confusion matrix, i.e. the percentage of systems in each class that are either classified correctly (along the diagonal) or
 as objects of other classes (off-diagonal entries). For the sake of completeness, we list the results both for an unbiased
 sample and one biased according to eq.~(\ref{eq:cuts}). The recognition of lensed quasars and quasar pairs does not vary appreciably,
 the main difference being the misclassification of bright single quasars as possible quasar pairs.

 To quantify the performance of the classifiers on the SQLS, we considered just those objects that are flagged as targets
 from the ANNs (Sect.\ref{sect:techn1}), since in a realistic search those are the systems that would be inspected with
 pixel-by-pixel techniques. The results are summarised in Table~\ref{tab:finalGBTs}. The importance of PCA versus KPCA is secondary,
 although the GBTs trained on KPCA use a smaller number of components and produce smoother classification probabilities,
 which are then more reliable for prediction purposes.

\subsubsection{Other Classification Models}
\label{sect:others}
We also investigated using the first 500 PCs instead of the first 200 KPCs, but the KPCs sill gave better accuracy.
 In addition, we also investigated the performance of logistic regression, a single deep decision tree, support vector machines,
 random forests, and neural networks for classification using the first 200 KPCs. As with the gradient boosted decision tree
 classifier, the tuning parameters for each classifier were chosen with cross-validation, with the exception of the neural network.
 We chose the tuning parameters of neural network after some initial exploration and employed
 early stopping with a validating set of $25\%$ of the training data. Although gradient boosting gave the best performance,
 the theoretical performance of support vector machines, random forests, and neural networks were all comparable to that. 

\section{Extensions to other regimes}
The choices made in the previous sections are not unique. For example, the target selection can be trained on selected samples
 as done for the candidate selection, although this does not bring to substantial differences in the performance of the learners.
 Also, the same techniques can be used to select targets in the deblended regime. For the sake of completeness, 
 in this Section we briefly discuss extensions of our techniques to those two aspects.

\subsection{Selection bias}

\begin{figure*}
 \centering
 \includegraphics[width=0.45\textwidth]{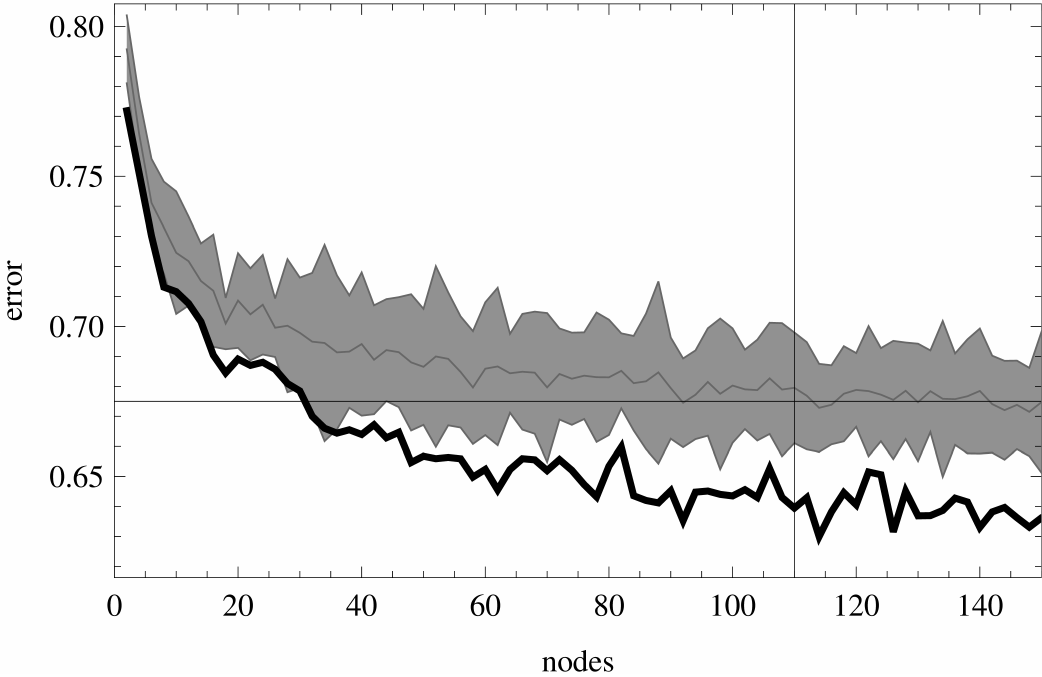}
 \includegraphics[width=0.45\textwidth]{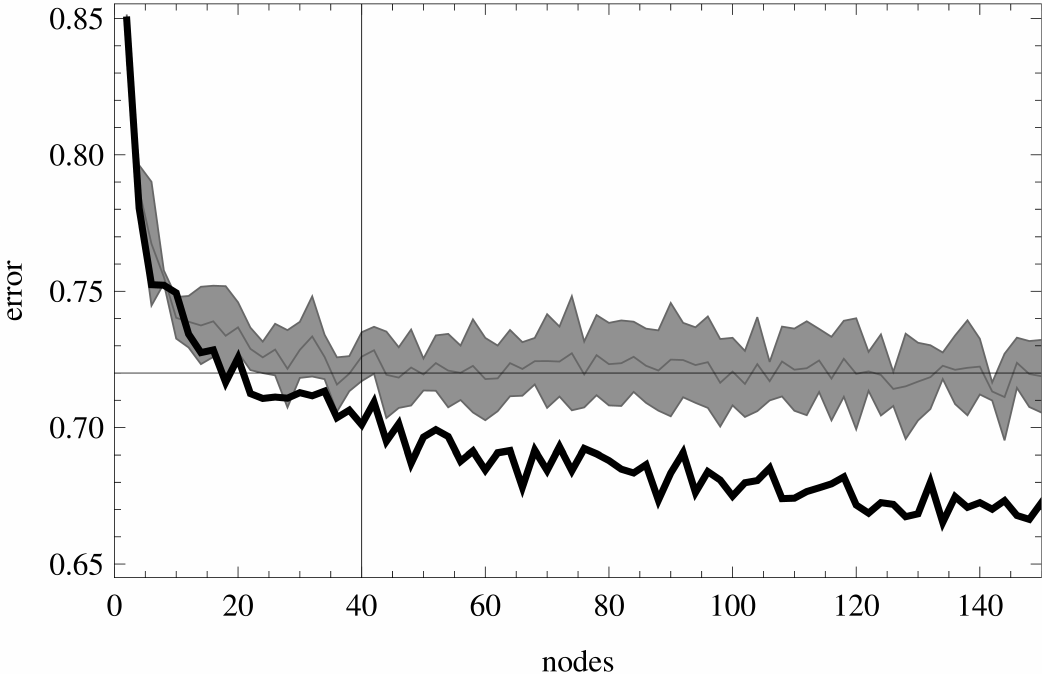}\\
 \includegraphics[width=0.45\textwidth]{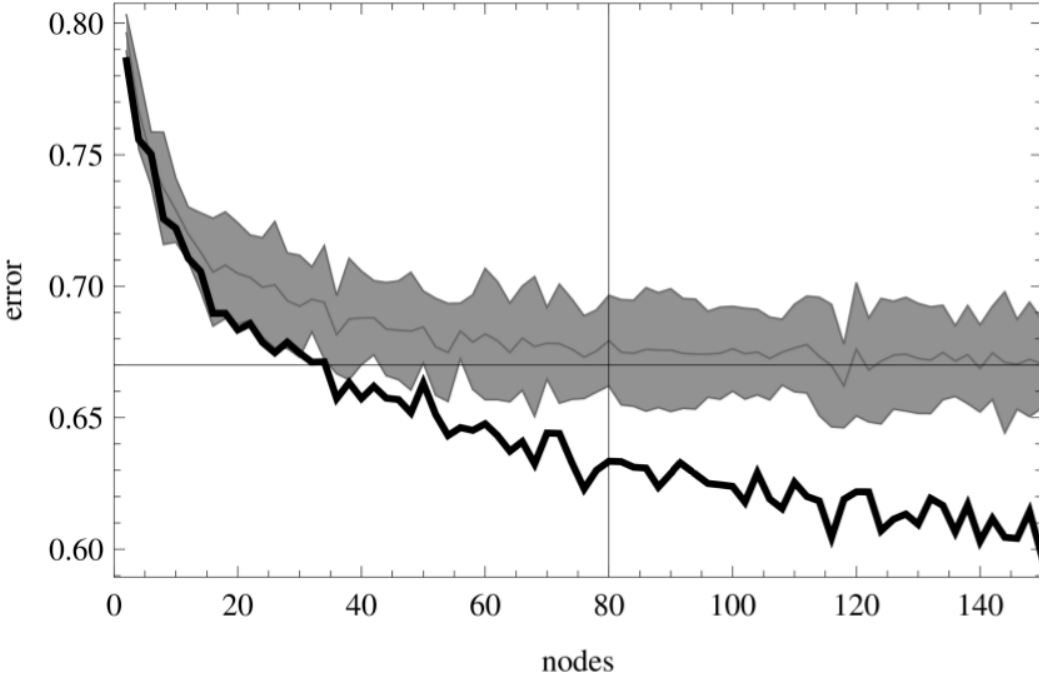}
 \includegraphics[width=0.45\textwidth]{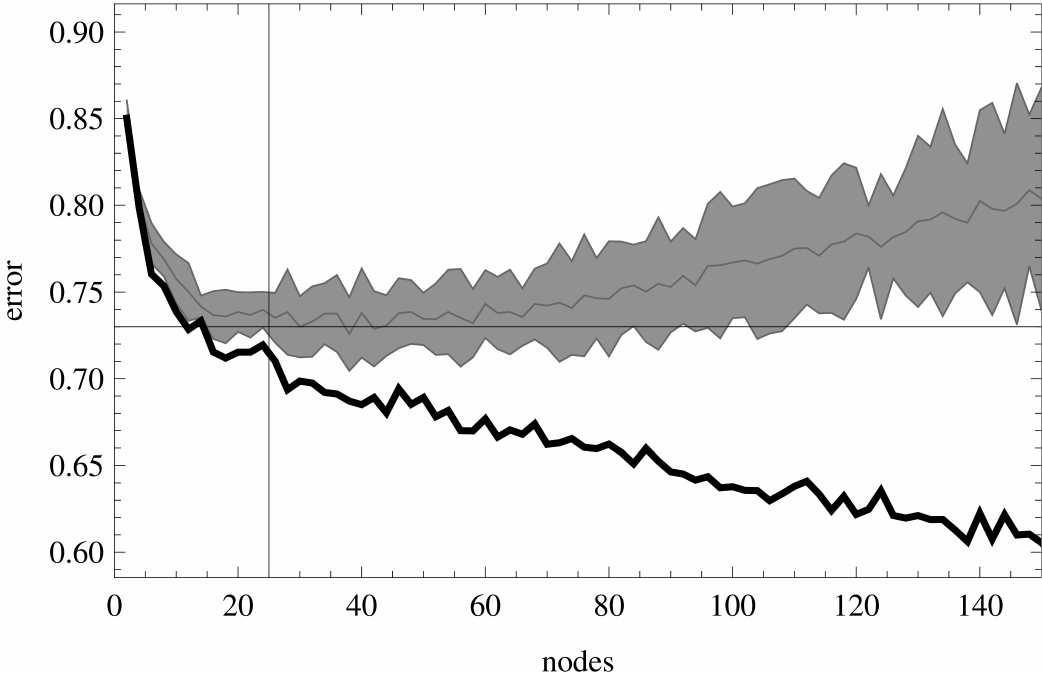}
 \caption{\small{Exploration of the ELMs performance on training and
 validating sets as a function of nodes. Full black lines: test-set
 error; grey stripes: average and dispersion of the validating-set
 error within $1\sigma.$ Top (resp.bottom) panels show the performance
 of ELMs on the $griz+W1+W2$ bands (resp.adding the second moments),
 without (left) or with (right) a heavy selection bias towards objects
 with the photometry of bright quasars. Axes cross at the best
 validation-set error and at the number of nodes that is required to
 attain it.}}
 \label{fig:ELMnodes}
\end{figure*}

The simplest searches for lensed targets exploit magnification bias,
since the brightest objects in a population are likely lensed. On the
other hand, this way just the stretched, bright tail of the QSO
population is considered, which is not representative of most of the
lensed quasars in a survey. Such an effect must be accounted for when
tailoring the data mining techniques to a particular survey, such as
the SQLS, which can be strongly biased. In the next Section, we will
do so \textit{ex post} by suitably interpreting the membership
probabilities that are predicted by the learners, where this is
possible. Here, we illustrate the effect of a strong selection
bias on the theoretical performance and demands of the learners.

We have simulated different sets of objects, retaining just those
systems that satisfy the cuts of eq.~\ref{eq:cuts}.
For exploratory purposes, we have simply relied on a set of ELMs (Sect.~\ref{sect:techn1})
 to train the target selection.
 We varied the number of nodes $M$ between 2 and
150. For each choice of $M,$ we have drawn random values of the hidden
weights $\boldsymbol{\alpha}$ a hundred times, solved for the output
weights $\boldsymbol{\beta}$ on the test set for each of those
realizations, evaluated $R_{err}$ on five validation sets and selected
just the $(\boldsymbol{\alpha},\boldsymbol{\beta})$ that minimize the
average error, which simply defined as $\sqrt{R_{err}}$ on every
validation set. This mimics the early stopping criterion of ANNs,
while training just on the output weights. The features are
standardised and the entries of each $\boldsymbol{\alpha}$ are of the
kind $\alpha_{m,l}=\arcsin^{3}(u),$ where $u$ is drawn uniformly in
$[0,1].$ This ensures that most of the separating hyperplanes in
feature space pass through the bulk of the dataset, while also
covering the tails of the distribution.
 
The results are shown in Figure~\ref{fig:ELMnodes}, in four cases:
test and validation sets drawn as in the rest of this work or with a
heavier selection bias in optical/IR bands (equation~\ref{eq:cuts}),
considering just the magnitudes  ($p=6$ features per object) or also
the second moments ($p=13$). If a dataset is unbiased, it offers a
complete description of a population, so that most objects will occupy
different regions of feature space and the classifiers manage to
separate most of them, which is reflected in the smaller
validation-set error for the unbiased simulated sample. On the other
hand, learners that are trained on heavily biased samples are tailored
on simple and lean catalog searches, where they can repay their modest
theoretical performance. The information carried by the morphological
parameters is not always useful to the learners, especially in the
biased case. In fact, the average distance between objects increases
rapidly as the dimensionality of feature space is increased, making
the data sets too sparse and stretching the differences between
objects in the training and validating sets to fictitious levels.

\subsection{Larger separations or better image quality}
\label{sect:deblended}

\begin{figure}
 \centering
 \includegraphics[width=0.45\textwidth]{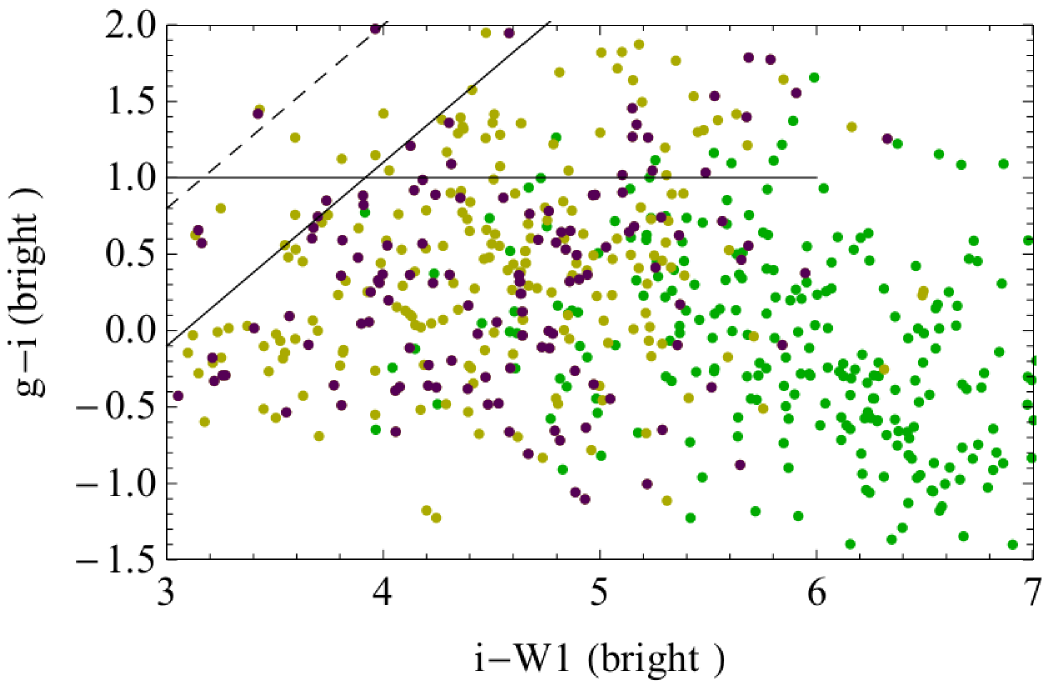}\\
 \includegraphics[width=0.45\textwidth]{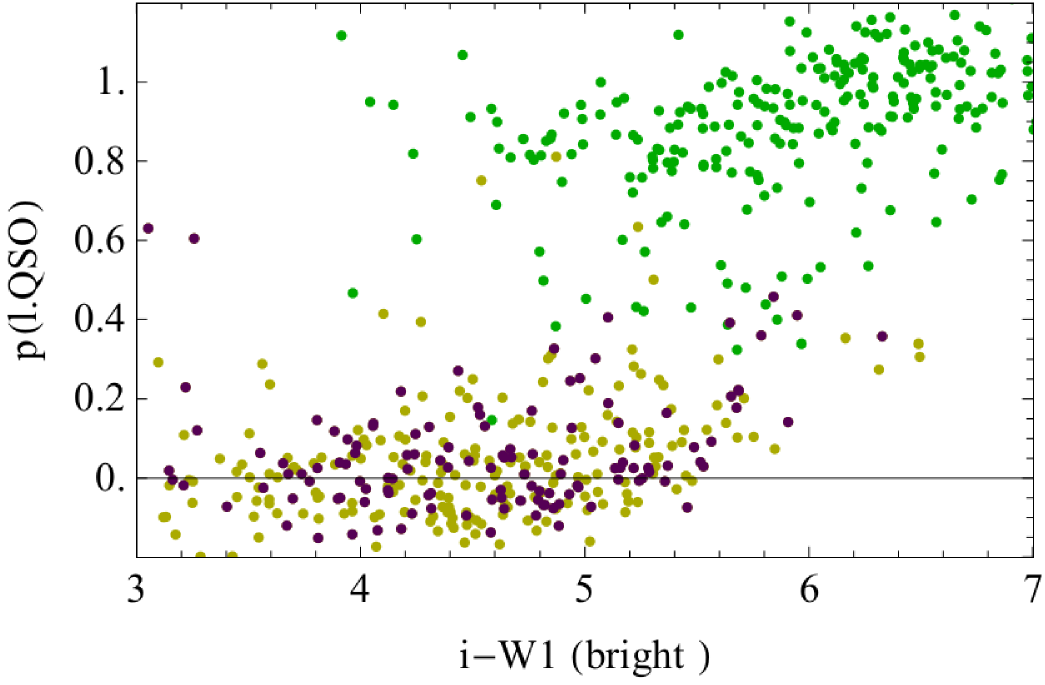}\\
 \includegraphics[width=0.45\textwidth]{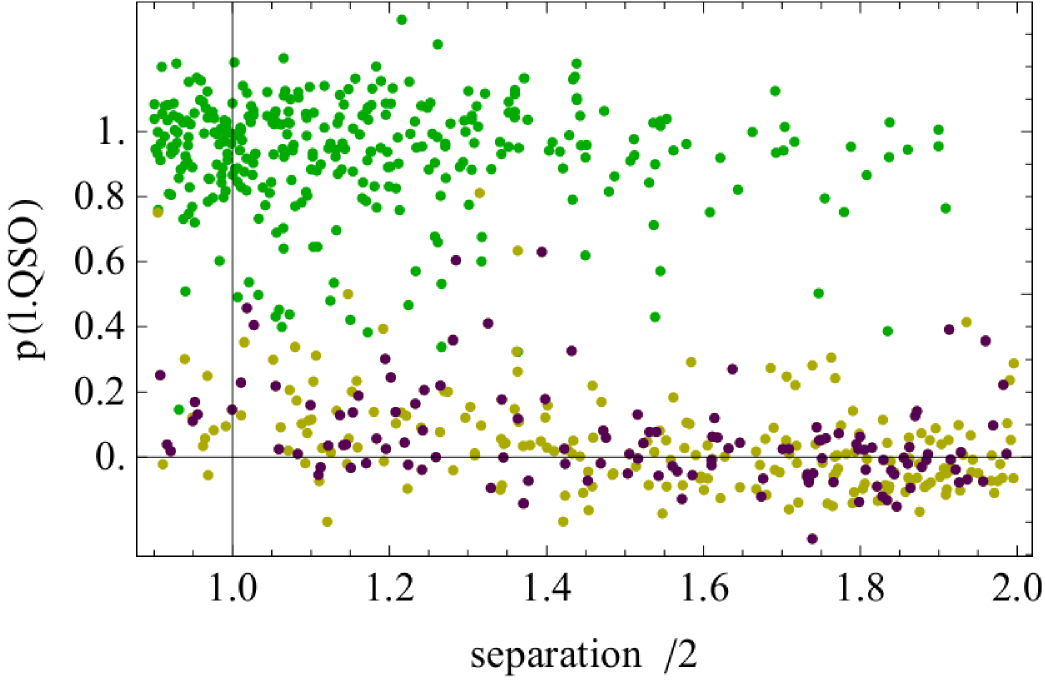}
 \caption{\small{Simulated systems in the deblended regime, with dark
 green (resp.yellow, dark red) representing \textit{l.QSO} (resp.
 {2QSO}, {ndd}) objects. Top: $g-i$ and $i-W1$ colours
 of the brightest QSO image. Middle: output $p(l.QSO)$ versus $i-W1$
 colour of the bright image. Bottom: output $p(l.QSO)$ versus half of the image-separation in arcseconds.
 A depth $i=24.0$ and PSF FWHM$=0.85''$ have been adopted here. }}
\label{fig:deblended}
\end{figure}
\begin{figure}
 \centering
 \includegraphics[width=0.45\textwidth]{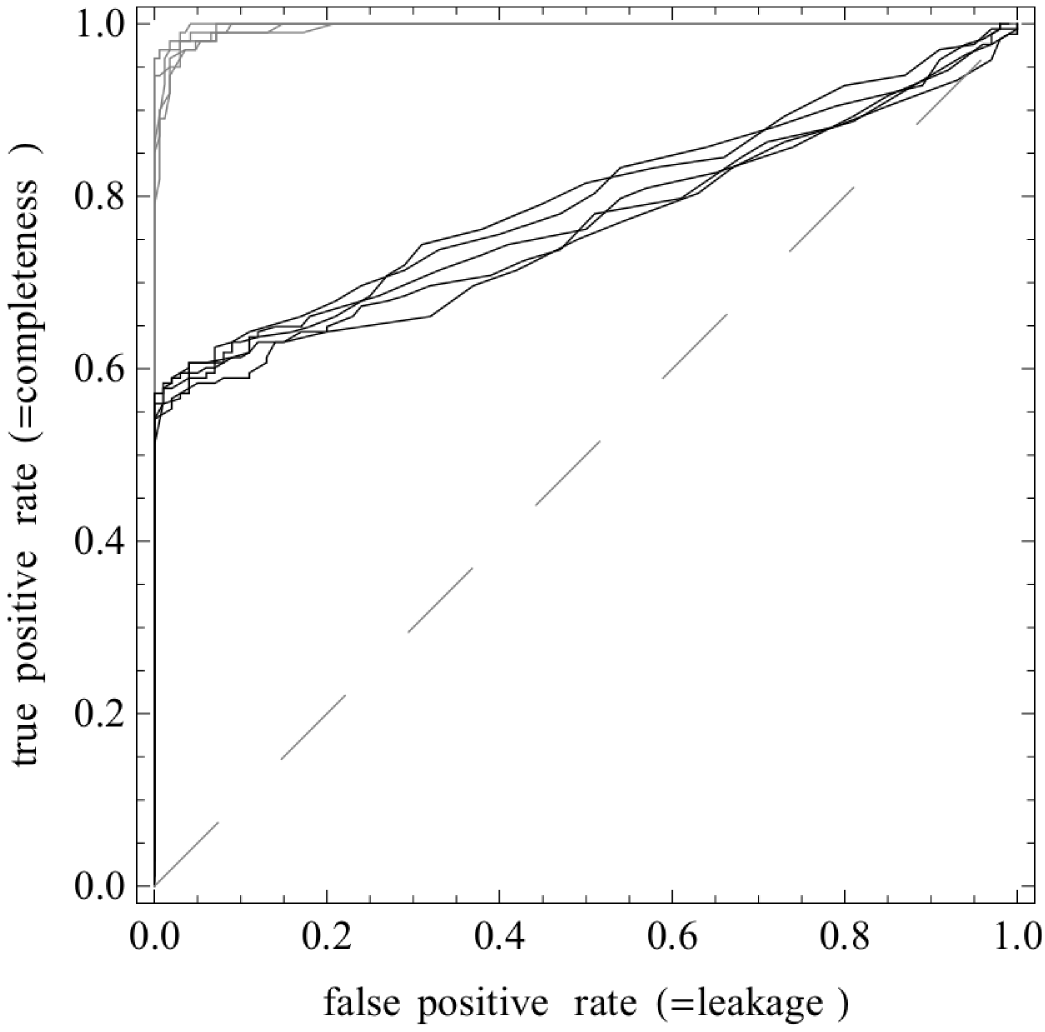}\\
 \caption{\small{ROC plots for target selection in the deblended regime. The
  grey lines show the performance on the five validating sets when just $l.QSO$
  systems are considered as true positives. The black lines refer to the selection of
  both $l.QSO$ and $nlo$ systems. As $nlo$ systems have a faint deflector,
  they are mostly classified as QSO pairs.}}
\label{fig:debROC}
\end{figure}

Depending on the image separation and imaging quality of a survey, an
appreciable fraction of lensed quasars can appear as close
($\theta_{sep}\leq4''$) QSO pairs. This can be the case, for example,
with the claimed depth ($i\approx24$) and image quality (median
FWHM$\approx 0.85$) of DES and will be even dominant for Gaia
(resolution $\approx 0.15''$). Here we illustrate how the same
techniques, specifically ELMs (Sect.s \ref{sect:techn1},
\ref{sect:annelm}), can be used to study the search for lensed quasars
in the deblended regime.

We have simulated two classes of objects, lensed quasars (l.QSO) and
line-of-sight quasar pairs (2QSO), with the same procedures as for the
blended case, plus a third class that will be described below. As the
fainter image of a lensed QSO is also closer to the deflector, one
must add a differential reddening between multiple images, since
colour comparison is a criterion for colour selection of close-by,
quasar-like objects. We refer to \citet{ogu06} for more detail on the
acceptable region for the reddening vector in $griz$ bands. Also, the
PSF-photometry and morphology of the faintest QSO image are more
contaminated by the deflector's flux. Still, we can safely suppose
that the QSO images will not appear as extended, as is confirmed in
practice by previous searches.

For these reasons, the data mining in the deblended regime is trained
on the following features: PSF magnitudes in $griz$ of the images;
overall $W1,W2$ magnitudes; flattening and position angle of the faint
image in $griz$ bands; and faint-to-bright image position angle. When
simulating l.QSO systems, we compare the overall $i-$band magnitude of
the simulated cutout with the global PSF magnitudes of the QSO images
to estimate the mean surface-brightness $SB(R_{E})$ of the deflector
within the Einstein radius. The distribution of $SB(R_{E})$ shows a
secondary peak beyond $\approx 18.5,$  so if a simulated object has
$SB(R_{E})>18.5,$ we store it in a third class ({ndd}, no
detected deflector).

The results of this procedure are displayed in
Figure~\ref{fig:deblended}, for the test set only for visual
convenience. We have chosen the half-separation in the last panel because that is
also a rough estimate of the Einstein radius for the $l.QSO$ systems.
 If the $g-i$ and $i-W1$ colours of the bright QSO image
are considered, regardless of the class, the $i-W1$ colour is larger
than in equation~\ref{eq:cuts}, simply because the $W1$ magnitude
encloses the flux from the whole system, because of the large WISE
FWHM. The full lines in the top panel show the colour cuts as in
eq.~(\ref{eq:cuts}), the dashed line is simply shifted by
$1.2\times2.5\log_{10}(2)\approx 0.9,$ as would be expected in a QSO
pair with comparable $i-$band fluxes between the two objects. The
middle panel shows the output $p(l.QSO)$ as a function of bright image
$i-W1,$ whereas in the bottom panel $p(l.QSO)$ is examined against the
image separation. It becomes evident that, even if 2QSO systems become
more frequent and dominate at larger separations, data mining on the
photo-morphological features is still effective at separating the
classes, except for those few systems where the deflector is not
bright enough -- as exemplified by the {ndd} systems, which are
unrecognised lensed quasars.

A more quantitative analysis is shown in the ROC plots of Figure \ref{fig:debROC} on the validating sets.
 If just $l.QSO$ systems are our main interest (grey curves), the performance in excellent. The recovery of both
 $l.QSO$ and $nlo$ systems (i.e. regardless of the deflector's brightness, black curves)
 is less sharp, as $nlo$ systems are mostly classified as $2QSO.$

\section{Conclusions}

We have applied machine-learning techniques to the search for
gravitationally lensed quasars in wide field imaging surveys, breaking
down the problem into two stages, target selection and candidate
selection. In the target selection stage, promising systems were
selected based solely on information available at the catalogue level.
Focusing on SDSS and WISE data, in order to compare with the actual
output of the SDSS Quasar Lens Search, we used thirteen parameters: the
magnitudes in $griz$ and WISE $W1,W2$ bands, and the axis ratios and
position angles in the SDSS $griz$ bands. In the candidate selection
stage, we returned to the images of the targets in order to narrow
down the search: we used 10 arcsecond ($25\times25$ pixel) SDSS cutout
images in $griz$, and reduced the dimensionality of the feature space
(from 2500 to 200) via kernel-PCA.

In order to have a large set of objects, for training and validation
we used a mixture of simulated and real objects, all assuming or
directly sampling the SDSS imaging conditions. In particular, we
simulated systems in three classes (lensed QSO, QSO+LRG alignment,
QSO+QSO alignment), and added a sample of Blue Cloud (BC) galaxies
drawn from SDSS. 
For the candidate selection stage, we discarded
the BC galaxies and included a simulated sample of single quasars.
This allowed us to explore the range of true- and false-positives in a
strategy that might be followed in an SDSS lensed quasar search
restricted not to use any spectroscopic information.

Artificial Neural Networks were used to separate lensed quasar targets
from false positives. In particular, we focused on single-hidden
layer, feed-forward networks, which are trained with backpropagation
and early stopping. The use of such ANNs on the photo-morphological
features enables the separation of QSO-like objects from Blue-Cloud
galaxies, which would otherwise be a dominant source of contamination
for samples of extended objects selected in $griz+W1+W2$ bands. In
particular, with hard cuts in optical/IR bands (eq.\ref{eq:cuts}) and
the requirement of extended morphology, about a tenth of the Blue
Cloud galaxies leak into the sample of targets with extended
morphology brighter than 19 in $i-$band. Use of ANNs prunes these away
effectively, reducing the leakage to the percent level. When tested on
the SQLS morphologically-selected sample, which is biased towards
bright quasars, the best ANNs give a twofold (up to threefold)
increase in purity at the price of a $10\%$ (or $20\%$) reduction of
the completeness. 

Returning to the pixels for candidate selection, we trained
Gradient-Boosted Trees on the kernel-PCA features extracted from 
simulated images of plausible targets, whose photometry obeys a
selection bias similar to that of SQLS objects
(equation~\ref{eq:cuts}). When tested on the targets selected by the
ANNs, this system gave a final purity of $70\%$ while correctly
identifying $70\%$ of the true positives.
% PJM: I dont understand these numbers. The table says 70 -70!!
%  - i.e. final completeness
% between $56\%$ and $63\%.$

In the broader context, our novel technique is highly complementary to
and synergistic with alternative approaches that are or have been
proposed. For example the ANN catalog level selection, {with its
high completeness and reasonable purity}, could be used to pre-select
targets for human classifiers in a citizen science project, or to be
fed to robots designed to automatically model the lensing features in
pixel space \citep[e.g.][]{mar09}. These approaches could be run in
parallel to the kPCA method proposed here, and with each other: in
order to find reliably large samples of lensed quasars it is possible
that many of these techniques will have to be used in parallel. 

{A significant advantage of machine learning techniques, like those
explored here, is that they are very fast both to train and to run as 
classifers. Speed will be essential to find large sample of lensed
quasars in ongoing and upcoming imaging surveys, such as PS1, DES,
HSC, and LSST. Searches in these surveys that use traditional lens
finding techniques would require several seconds of CPU or investigator
time per system for very large numbers of objects.
 In contrast,  with a run-time of $\approx10^{-3}$ seconds
per system, the machine learning target selection presented here could
perform a catalog search over the DES-wide footprint in just a few
hours on a 12 core desktop workstation.} Assuming
conservatively a purity between 20\% and 50\% from this catalog
search, finding the brightest 100 lenses in DES would require running
the pixel based algorithm on only a few hundred cutouts, which is a
trivial computational task.

As follow-up efforts provide larger and larger samples of false and
true positives for each survey, they will provide new training sets to
improve algorithms such as these. Furthermore, as data are reprocessed
and improved the search can also be repeated with minimal expense.
Finally, these techniques are inherently repeatable so that the
results can be reproduced by independent users. 
{What deserves further investigation is how well machine classifiers
 perform when presented with test sets that contain unusual
lenses. An example in the present context might be a system with 
 significant differential reddening or microlensing.
We do not expect the extension of the methods applied here to larger
feature sets to pose significant problems: additional information,
particularly from the time domain \citep[e.g.][]{Schmidt10} would be
straightforward to include. Including catalog variability parameters,
or time series of image cutouts, would be an interesting next step.}

In conclusion, we have used the SQLS and simulated samples to
illustrate the power of machine learning techniques in finding
gravitationally lensed quasars.  We have illustrated how it works on
blended (and hence difficult) objects, but it can be naturally applied
also to the deblended regime (Section~\ref{sect:deblended}). Even
though these techniques might seem cumbersome (perhaps even
``Rococo''), it is striking how much improvement is afforded over
simpler, traditional techniques. For example, the ANN catalog-level
search yields a purity up to $\approx 60\%$, which is appreciably better
than what was achieved by SQLS and an order of magnitude higher than
what can be achieved with simple colour cuts in $griz$+WISE and the
requirement of spatial extension. This means reducing the follow-up
effort by the same amount, a crucial goal if one wants to confirm
large samples of lenses. For example, in order to find of order 100
lensed quasars from a Stage III experiment like DES one would have to
follow-up only $\approx$200 candidates, as opposed with the thousands
required for purities below 10\%. With the addition of time domain
information, the methods should further improve their performance in
terms of purity, which will be key to containing follow-up costs.

\section*{Acknowledgments}

AA, TT acknowledge support from NSF grant AST-1450141 ``Collaborative
Research: Accurate cosmology with strong gravitational lens time
delays''. AA, BCK, and TT gratefully acknowledge support by the
Packard Foundation through a Packard Research Fellowship to TT.  The
work of PJM was supported by the U.S.  Department of Energy under
contract number DE-AC02-76SF00515. The authors are grateful to Robert
Brunner and their friends and collaborators in the STRIDES project
(especially Matt Auger, Chris Kochanek, Richard McMahon, and Fernanda
Ostrovski) for many useful suggestions and comments about this
work. The \citet{om10} catalog is freely available at
\\ {\tt https://github.com/drphilmarshall/OM10}\\ 
STRIDES is a broad external collaboration of the Dark Energy Survey,
\\ {\tt http://strides.physics.ucsb.edu}\\

\onecolumn
\appendix
\section{Simulated galaxies and quasars}
\begin{figure}
 \centering
 \includegraphics[width=0.6\textwidth]{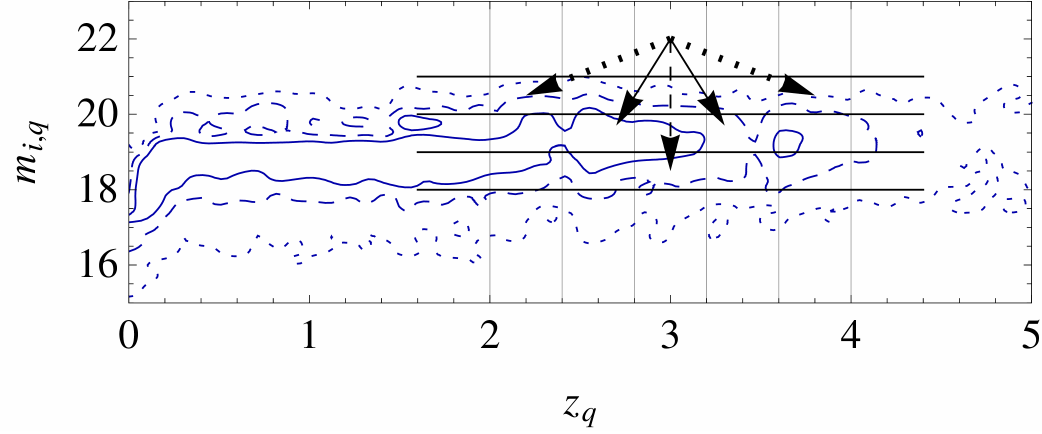}
\caption{\small{Schema of the empirical matching procedure, in this case for the quasars.
 The primary observables are binned; the conditional probability $\theta(\mathbf{r}|\mathbf{p})$ is constructed
 in each bin; when a new system is considered, the conditional distribution of its observables is interpolated
 among its nearest neighbours in the space of primary observables.}}
\label{fig:sparseinterp}
\end{figure}
For the simulated systems for this work, we needed the magnitudes of QSOs and LRG in $griz$ and $W1,W2$ bands,
 the effective radii of the LRGs and a prescription to link those to the primary observables in eq.s (\ref{eq:lrg}) and (\ref{eq:qso}).
 We proceed by assembling a catalogue of QSOs and one of LRGs (spectroscopically-confirmed) from SDSS and WISE. For each
 of the quasars, we retrieve the redshift $z_{q}$ and apparent magnitudes in $(g,r,i,z,W1,W2)$ optical/IR bands. For each
 LRG, we retrieve its redshift $z_{g},$ magnitudes in optical/IR bands, velocity dispersion $\sigma$ and $r-$band effective radius
 $R_e.$ Both the QSO and LRG queries are split into redshift bins of width $\delta z=0.1,$ with $10^3$ objects in each bin,
 so as to ensure an even coverage of the redshift range.

From the SDSS+WISE catalogues we can build the conditional probability $\theta(\mathrm{r}|\mathrm{p})$
 that a QSO (resp.LRG) with observables $\mathrm{p}=(z_{q},m_{i})$ (resp. $z_{g},\sigma$) has some values
 of the remaining observables $\mathrm{r}.$
 We first bin the QSO (resp.LRG) catalogue in redshift and $m_{i}$ (resp. $\sigma$), so that in each bin the mean and covariance
 of the remaining observables can be easily computed. In other words, we have a characterization of the conditional probability at
 some discrete locations, $\theta(\mathbf{r}|\mathbf{p}_{l})_{l=1,...,N_{bins}}.$
 The next step is to build a smooth interpolation of those, across the whole range of redshift and $m_{i}$ (or $\sigma$).
 This is needed in order to assign values of $\mathbf{r}$ also to systems whose primary observables $\mathbf{p}$ are not
 well represented by the SDSS+WISE catalogue. Figure \ref{fig:sparseinterp} summarizes the main steps.
 
Given a pair $\mathbf{p}=(z_{q},m_{i})$ for a QSO in the simulated catalogue, we find its corresponding bin,
 say $\mathbf{p}_{n,m}=(z_{q,n},m_{i,m}),$ and build a sparse interpolation for $\theta(\mathbf{r}|\mathbf{p})$ as
\begin{equation}
\theta(\mathbf{r}|\mathbf{p})=
\frac{ \sum\limits_{i,j}\theta(\mathbf{r}|\mathbf{p}_{i,j})\exp[-|i-n|-|j-m|] }{ \sum\limits_{i,j}\exp[-|i-n|-|j-m|]  }\ .
\label{eq:spint}
\end{equation}
The fast exponential fall-off of the weights ensures that just the nearest (populated) neighbour bins give a contribution to the
 interpolated probability. This is useful in order to smooth out bin-to-bin noise and obtain a smooth probability also for bins
 that are not well populated.
 In practice, instead of computing the whole probability distribution in each bin,
 we compute the mean and dispersions of $\mathrm{r}$ and interpolate those similarly to eq.~(\ref{eq:spint}).
 The procedure is the same for LRGs, with the obvious changes, and allows us to draw magnitudes (and effective radii)
 that are as close as possible to the ones found in SDSS and WISE.

\begin{figure}
 \centering
 \includegraphics[width=0.45\textwidth]{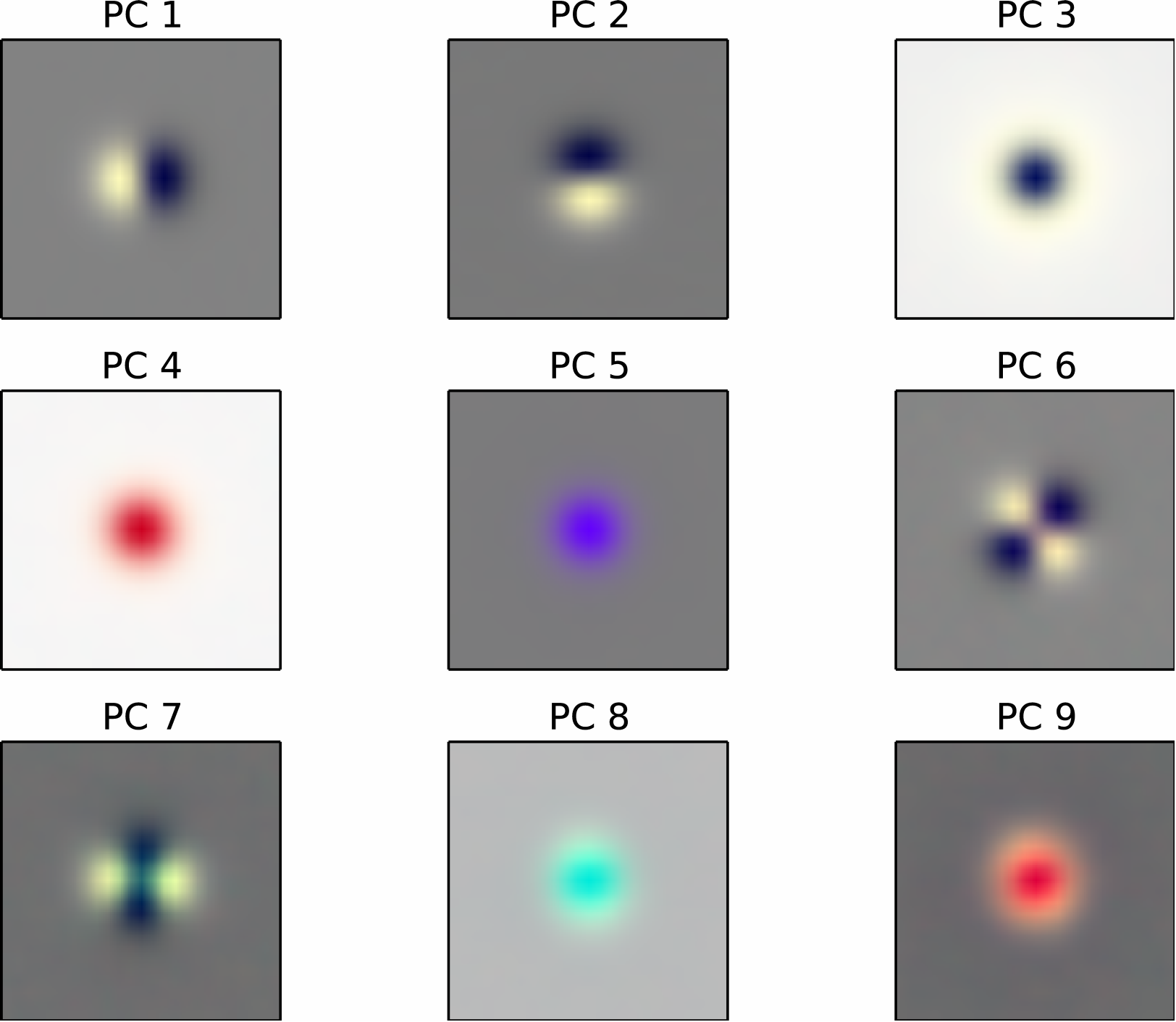}
 \includegraphics[width=0.45\textwidth]{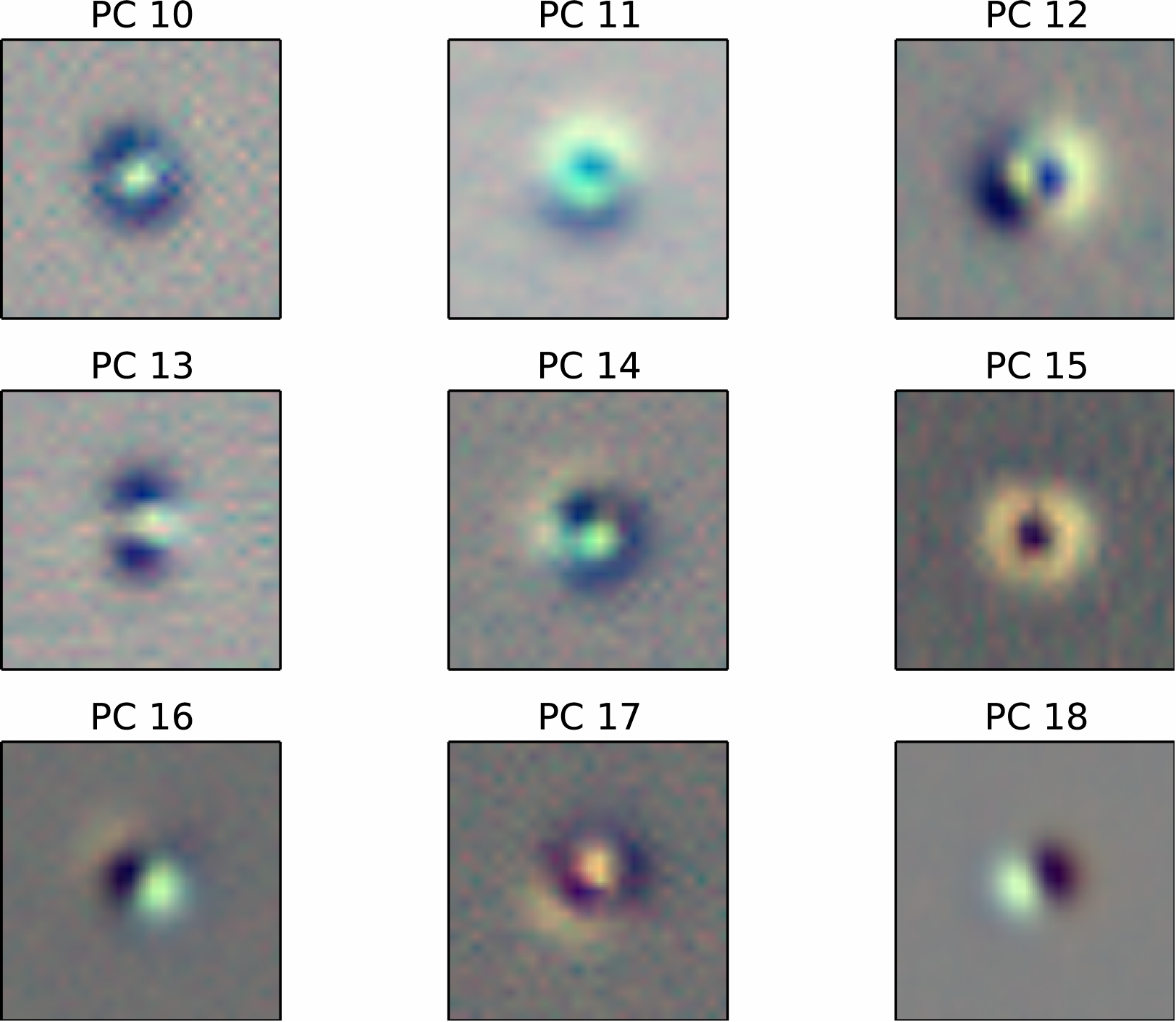}
\caption{\small{Composite $gri$ plots of the first 18 principal components of the simulated sample.}}
\label{fig:pcapics}
\end{figure}
\section{Dimensional reduction and Machine-learning algorithms}
In all of the machine-learning techniques the features are standardized, i.e. we subtract the test-set average and
 divide by the variance. This way, we operate on feature vectors with entries typically within $[-1,1],$ overcoming
 issues related to the choice of units of measure and zero-points.

\subsection{PCA and kPCA}
When separating objects in different classes, we tacitly suppose that their features are linked to one another through relations
 that, in general, must be found. In other words, the data features are usually correlated and we may seek new combinations of features
 that naturally follow these correlations. In Principal Component Analysis (PCA), a training set of vectors
 $\left\{\mathbf{x}_{i}\right\}_{i=1,...,N}$ is linearly transformed into a set $\left\{\mathbf{f}_{i}\right\}_{i=1,...,N}$
 whose new features are uncorrelated, i.e. $\langle f_{n}f_{m}\rangle=(1/N)\sum_{i}f_{i,n}f_{i,m}=0.$ This is
 simply achieved by diagonalizing the data covariance matrix $C_{n,m}=\langle x_{n}x_{m}\rangle.$ The principal components
 are the eigenvectors $\mathbf{w}_{1},...,\mathbf{w}_{p}$ of $C_{n,m},$ and the coordinates of a vector $\mathbf{x}_{i}$
 in this basis are the projections $t_{i,r}=\mathbf{w}_{r}\cdot\mathbf{x}_{i}.$
 
The sum
\begin{equation}
var(r)=\frac{\sum_{m=1}^{r}\langle t_{m}^{2}\rangle}{\sum_{m=1}^{p}\langle t_{m}^{2}\rangle}
\end{equation}
is the {fraction of explained variance} within the dataset up to the $r-$th component. Because of this, PCA can also be
 used to de-noise and reduce the dimensionality of the problem at hand, by retaining just the projections onto the first $r$ components,
 at the price of encompassing just a fraction $var(r)$ of the whole feature variability.
 
In our case, each feature vector consists of four concatenated $25\times25$ pixel cutouts, one per band, normalized to the total flux.
 Hence, the principal components are combinations of shapelet images in the four $griz$ bands. Figure \ref{fig:pcapics} shows
 $gri$ composites of the first $18$ principal components for our training set of simulated cutouts.
 
%\textbf{AA: Add kPCA, two formulae and quick justification through Mercer's theorem.}
Kernel Principal Component Analysis \citep[kPCA,][]{Scholkopf1998a} provides a very elegant extention of the PCA approach,
 by supposing that the feature space is a (perhaps unfortunate)
 projection of a $p-$dimensional manifold which resides in a higher-dimensional feature space. In fact, in PCA-based classification
 one relies just on the coordinates $t_{i,r}=\mathbf{w}_{r}\cdot\mathbf{x}_{i}$ rather than the principal component vectors themselves.
 Within kPCA, the scalar product $\mathbf{x}\cdot\mathbf{w}$ in $\mathbb{R}^{p}$ is replaced with a semi-positive definite kernel
 $k:\mathbb{R}^{p}\times\mathbb{R}^{p}\rightarrow \mathbb{R},$ provided there is a nonlinear map
 $\Phi:\mathbb{R}^{p}\rightarrow\mathcal{H}$ onto a higher-dimensional Hilbert space, such that
 $k(\mathbf{x},\mathbf{y})=\langle\Phi(\mathbf{x}),\Phi(\mathbf{y})\rangle$ is a scalar product in $\mathcal{H}.$
 Now the task it so diagonalize the new covariance matrix operator\footnote{Here the `dagger' apex denotes the hermitian conjugate
 in $\mathcal{H},$ such that $\mathbf{v}^{\dag}\mathbf{u}=\langle\mathbf{v},\mathbf{u}\rangle\ .$}
 in $\mathcal{H}$
\begin{equation}
\tilde{\mathrm{C}}=\frac{1}{N}\sum\limits_{i=1}^{N}\Phi(\mathbf{x}_{i})\Phi(\mathbf{x}_{i})^{\dag}
\end{equation}
The eigenvectors $\mathbf{v}_{r}$ of $\tilde{\mathrm{C}}$ must be linear combinations of the new feature vectors in $\mathcal{H},$
\begin{equation}
\mathbf{v}_{r}=\sum\limits_{i=1}^{N}a_{r,i}\Phi(\mathbf{x_{i}})\ ,
\end{equation}
 which gives a new eigenvalue equation for $K_{i,j}=(1/N)\langle\Phi(\mathbf{x}_{i}),\Phi(\mathbf{x}_{j})\rangle\equiv(1/N) k(\mathbf{x}_{i},\mathbf{x}_{j})$
 and the weights $a_{j,i}$
\begin{equation}
\mathrm{K}\mathbf{a}=\lambda\mathbf{a}\ .
\end{equation}
Once the (orthonormal) weight vectors $a_{r,i}$ are found, the kPCA components in $\mathcal{H}$
 of a feature vector $\mathbf{f}\in\mathbb{R}^{p}$ are simply
\begin{equation}
\tilde{t}_{r}=\ 
\langle \mathbf{v}_{r},\Phi(\mathbf{f})\rangle=\sum\limits_{i=1}^{N}a_{r,i}\langle\Phi(\mathbf{x}_{i}),\Phi(\mathbf{f})\rangle\ 
=\  \sum\limits_{i=1}^{N}a_{r,i}k(\mathbf{x}_{i},\mathbf{f})\ .
\label{eq:ktrick}
\end{equation}
 
 A suitable adaptation of Mercer's theorem \citep{mer19} ensures that $\Phi$ exists whenever $k$ is semi-positive definite,
 however it will generally map $\mathbb{R}^{p}$ into an infinite-dimensional Hilbert space. Fortunately, since just the
 kPCA components are used for classification, we never need to compute the map $\Phi$ explicitly and the non-linear
 structure of the data is simply encoded in the kernel $k$ via eq.(\ref{eq:ktrick}).

\subsection{Artificial Neural Networks and Extreme Learning Machines}
\label{sect:annelm}
%
%\textbf{AA: 
% For a  totally random classifier, one would have $f_{i,k}=1/K$ and so $R_{err}=(K-1)/K.$}
Let us consider a smooth, increasing {activation function} $g:\mathbb{R}\rightarrow\mathbb{R}$ such that
\begin{equation}
\lim\limits_{x\rightarrow+\infty}g(x)=1\ ,\ \lim\limits_{x\rightarrow-\infty}g(x)=0\ .
\end{equation}
The idea underlying ANNs is to use $g$ to construct arbitrarily good approximations to given functions over the feature space
 $\mathbf{x}\in\mathbb{R}^{p}.$
In particular, any piecewise continuous function $g:\mathbb{R}^{p}\rightarrow\mathbb{R},$
 defined on a compact subset of $\mathbb{R}^{p},$ can be approximated by combinations of the kind
\begin{equation}
t=\sum\limits_{m=1}^{M}\beta_{m}g(\boldsymbol{\alpha}_{m}\cdot\mathbf{x}+a_{0,m})\ .
\label{eq:elm}
\end{equation}
The {number of nodes} M depends just on the tolerance that is desired in order to fit $y.$
 The same holds, quite naturally, for multidimensional (piecewise continuous) functions $\mathbf{y}=(y_{1},...,y_{K}),$
 mapping a compact subset of $\mathbb{R}^{p}$ into $\mathbb{R}^{K}.$ This is the case of classification problems,
 where $\mathbf{y}$ gives the membership probabilities to different classes for an object in feature space.
 A common choice for the activation function is the sigmoid $g(x)=1/(1+\mathrm{e}^{-x}).$

Approximations as in eq.(\ref{eq:elm}) are the core of {Extreme Learning Machines} \citep[ELMs,][]{elms},
  where the weights and biases  $(\boldsymbol{\alpha}_{m},a_{0,m})$ are held fixed and the parameters
  $\beta_{m}$ are adjusted. Specifically, operating a test set
  $\left\{\mathbf{x}_{i}\right\}_{i=1,...,N_{t}}$ with probability vectors $\left\{\mathbf{y}_{i}\right\}_{i=1,...,N_{t}},$
  the weights $\boldsymbol{\beta}_{m}$ can be found as:
\begin{equation}
\boldsymbol{\beta}_{m}=W^{\dag}_{i,m}\mathbf{y}_{i}\ ,
\label{eq:elm}
\end{equation}
 where $W^{\dag}$ is the pseudo-inverse of a matrix $W$ with entries
 $W_{i,j}=g(\boldsymbol{\alpha}_{j}\cdot\mathbf{j}+a_{0,j}).$
 We can also add a constant vector $\beta_{0}$ in the approximation $\mathbf{t}$, which is equivalent to having
 (at least) one of the activation functions $g(\boldsymbol{\alpha}_{j}\cdot\mathbf{x}+a_{0,j})$ equal to one. Similarly, we can
 regard the coefficients $a_{m,0}$ as part of the weight vectors $\boldsymbol{\alpha}_{m},$ if we embed $\mathbb{R}^{p}$
 in $\mathbb{R}^{p+1}$ as $(x_{1},...,x_{p})\mapsto(x_{1},...,x_{p},1).$ This is the convention that we will adopt in what follows.
 
The approximating solutions $\mathbf{t}(\mathbf{x})=(t_{1},...,t_{k})$ from ELMs are not necessarily probability vectors,
 which are required to have positive entries that sum to unity. Hence, when using ANNs for classification a final transformation
 $t_k\mapsto g_{k}(\mathbf{t})$ is made, with $g_{k}$ commonly chosen as the {soft-max}
\begin{equation}
g_{k}(\mathbf{t})=\frac{\exp[t_{k}]}{\exp[t_{1}]+...+\exp[t_{K}]}\ .
\end{equation}

The ANNs are trained by minimizing a loss function, wich can be either $R_{err}+R_{reg}$ or $R_{dev}+R_{reg},$
 which can be simply implemented by means of steepest descent methods since the derivatives with respect to the
 weights $(\boldsymbol{\beta},\boldsymbol{\alpha})$ can be computed analytically:
\begin{eqnarray}
R_{err}=\frac{1}{N}\sum_{i,k}(y_{i,k}-g_{k}(\mathbf{t}_{i}))^{2}\ \equiv\ \frac{1}{N}\sum_{i}R_{i}\\
\partial_{\beta_{m,k}}R_{i}=\frac{2}{N}(g_{k}(\mathbf{t}_{i})-y_{i,k})g^{\prime}_{k}(\mathbf{t}_{i})
g(\boldsymbol{\alpha}_{m}\cdot\mathbf{x}_{i})\ \equiv\ \delta_{k,i}g(\boldsymbol{\alpha}_{m}\cdot\mathbf{x}_{i})\\
\partial_{\alpha_{m,l}}R_{i}=\sum\limits_{k=1}^{K}\frac{2}{N}(g_{k}(\mathbf{t}_{i})-y_{i,k})g^{\prime}_{k}(\mathbf{t}_{i})
\beta_{k,m}g^{\prime}(\boldsymbol{\alpha}_{m}\cdot\mathbf{x}_{i})x_{i,l}\ \equiv\ s_{m,i}x_{l,i}\\
\end{eqnarray}
 Given the structure of the ANNs illustrated here, some convenient {back-propagation relations} hold among the coefficients,
 which descend from the additive nature of the loss function.
 For example, when the loss function is just $R_{err},$ one has
\begin{equation}
s_{m,i}=g^{\prime}(\boldsymbol{\alpha}_{m}\cdot\mathbf{x}_{i})\sum_{k=1}^{K}\beta_{k,m}\delta_{k,i}\ .
\end{equation}
 The gradients and back-propagation relations can be computed also for other choices of the loss function along these lines.

If a large number of nodes is used, or if the coefficients are completely unconstrained, one can {overfit}
 peculiar behaviours of observables in the test-set, which are not necessarily present in other datasets,
 thus losing predictive power. To avoid this, the loss function is also computed on some {validating} sets,
 with objects drawn from the same parent distributions as in the test set,
 and the optimization on the test set is stopped when the error on the validating sets does not decrease any more.
 For the analysis in Sect.\ref{sect:ANNtarg} we have assembled ten validating sets, as to have a characterization of the
 typical error and its variation over different datasets.
The addition of a regularization term has a similar effect. In fact, the classification problem corresponds to mapping the
 feature space into a classification manifold, parameterized by the membership probabilities $(y_{1},...,y_{K}),$ and
 regularization helps ensure that new objects will be mapped smoothly on the classification space.
Extreme Learning Machines offer an alternative to early stopping and regularization. First, the $\boldsymbol{\beta}$ coefficients
 from eq.(\ref{eq:elm}) have the smallest possible norm among all those that minimize the test-set error $R_{err},$ a similar outcome to
 regularization. Second, if a node has $\boldsymbol{\alpha}$ weights that are not well discriminatory, it will automatically
 have a small $\boldsymbol{\beta}$ coefficient, avoiding the need to back-propagate in $\boldsymbol{\alpha}.$
 Third, since $W^{\dag}$ in eq.(\ref{eq:elm}) is computed very fast, one can draw many random
 combinations of $\alpha$ weights and retain just the $\beta$ solution that minimizes the validation error
 -- it also minimizes the test-set error, by construction. 
 
\subsection{Gradient-Boosted Trees}
\label{sect:appGBTs}

For completeness, we summarize the gradient boosting algorithm for classification using decision trees. Our description follows that of \citet{Hastie2009a}, to which we refer the reader for further details. 

In gradient boosted decision trees the predicted output, $f(x)$, is modeled as a function of the inputs $x$ as a sum of $M$ trees
\begin{equation}
f(x) = \sum_{i=1}^M \nu T(x, \Theta_m).
\label{eq-sum_of_trees}
\end{equation}
The parameter $\nu$ is called the learning rate, and regularizes the model by controlling how much each tree contributes to the sum. A tree is formally expressed as
\begin{equation}
T(x, \Theta) = \sum_{j=1}^J \gamma_j I(x \in R_j),
\label{eq-tree}
\end{equation}
where $I(\cdot)$ denotes the indicator function that returns 1 if the argument is true, and 0 otherwise. The parameters $\Theta = \{R_j, \gamma_j\}$ represent the partition of the input space and the output for each partition, respectively, and are fit using a training set. The meta-parameter $J$ controls the number of partitions of the input space, and can be estimated through cross-validation, or using a separate validating set. From Equation (\ref{eq-tree}) one sees that a decision tree is a piecewise constant model.

Gradient boosting is an algorithm for minimizing a loss function that is modeled after techniques from numerical optimization. For classification, the functions $f_k(x)$ represent the unnormalized log-probability for the $k^{\rm th}$ class. They are related to the class probabilities $p_k$ by
\begin{equation}
p_k = \frac{\exp(f_k(x))}{\sum_{l=1}^K \exp(f_l(x))}.
\label{eq-classprobs}
\end{equation}
When there are more than two classes, gradient boosting fits a separate sum of trees to each class, and com,es them using Equation (\ref{eq-classprobs}). The training set outputs $y_1, \ldots, y_n$ are integer values representing the class labels, which take values from the set $\{G_1, \ldots, G_K\}$. 

The typical loss function for classification problems is the multinomial deviance
\begin{equation}
L(f) = -\sum_{i=1}^n \sum_{k=1}^K I(y_i = G_k) f_k(x_i) + \sum_{i=1}^n \log\left( \sum_{k=1}^K e^{f_k(x_i)} \right).
\label{eq-multinom_deviance}
\end{equation}
 (cf. eq.\ref{eq:Rdev}).
The optimization problem is to find the values of ${\bf f}_k = [f_k(x_1), \ldots, f_k(x_n)]^T$ that minimize Equation (\ref{eq-multinom_deviance}). Note that at this point it is not necessary that the functions $f_k(x)$ be a sum of trees. Sequential numerical optimization algorithms would express the solution as a sum of vectors
\begin{equation}
	\hat{\bf f}_{kM} = \sum_{m=0}^M {\bf h}_{km},
	\label{eq-optimization}
\end{equation}
where the first vector ${\bf h}_{k0}$ represent the initial guess, and the remaining $M$ vectors represent the updates as the algorithm evolves. It is common to choose the next vector in the minimization procedure as proportional to the gradient of the loss function, meaning that the algorithm evolves by stepping in the direction of largest negative change in the loss function. For example, using the initial guess ${\bf h}_{k0}$ the estimated function that minimizes the loss function would simply be $\hat{\bf f}_{k0} = {\bf h}_{k0}$. We can improve upon this estimate as
\begin{equation}
	\hat{\bf f}_{k1} = \hat{\bf f}_{k0} - \rho_1 {\bf g}_{k1},
	\label{eq-first_gbt_step}
\end{equation}
where $\rho_1$ represents the step size and ${\bf g}_1$ is the gradient vector evaluated at $\hat{\bf f}_{k0}$:
\begin{align}
	g_{ik1} =& \left. \frac{\partial L(y_i, f_k(x_i))}{\partial f_k(x_i)} \right|_{f_k(x_i) = \hat{f}_{k0}(x_i)} \nonumber \\
		    =& -I(y_i = G_k) + p_k(x_i).
	\label{eq-first_gradient}
\end{align}
The step size may be chosen to minimize the loss function $L(\hat{\bf f}_{k1})$. The procedure is repeated $M$ times, leading to Equation (\ref{eq-optimization}) where ${\bf h}_m = -\rho_m {\bf g}_{km}$. 

If we were only interested minimizing the loss on the training data, then we have everything we need in Equations (\ref{eq-optimization})--(\ref{eq-first_gradient}). However, we want to generalize our model to make new predictions, so we want to also minimize the loss function with respect to data outside of the training set. We could do this if we had the values of the gradient at other data points as well. The principal idea behind gradient boosting is to fit a tree to the negative gradient at each iteration, which then allows us to generalize the gradient beyond the training set.  In this case, the updates ${\bf h}_{km}$ are trees, leading to a solution (Eq. \ref{eq-optimization}) which is a sum of trees (Eq. \ref{eq-sum_of_trees}). The learning rate $\nu$ is analogous to the role of the step size $\rho_m$, as it controls how fast the optimization algorithm proceeds, and thus how much each tree contributes to the sum. Smaller values of $\nu$ ($\rho_m$) lead to better models (optimization solutions), but require more trees (optimization steps).

\label{lastpage}

\end{document}